\documentclass[twocolumn]{aastex631}
\usepackage{amsmath}
\usepackage{dcolumn}
\newsavebox\mysavebox
\newcommand{\dd}{\mathrm{d}}

\begin{document}
\title{Revisiting the role of the streaming instability for the cosmic-ray spectrum in the GeV to TeV range}
\author[0000-0002-5291-1645]{Linh Han Thanh}
\affiliation{Ruhr University Bochum, Faculty for Physics \& Astronomy, Theoretical Physics IV, Plasma-Astroparticle Physics, Bochum, Germany}
\affiliation{Ruhr Astroparticle and Plasma Physics Center (RAPP Center), Ruhr University Bochum, Germany}
\author[0000-0001-6692-6293]{Julien D\"orner}
\affiliation{Ruhr University Bochum, Faculty for Physics \& Astronomy, Theoretical Physics IV, Plasma-Astroparticle Physics, Bochum, Germany}
\affiliation{Ruhr Astroparticle and Plasma Physics Center (RAPP Center), Ruhr University Bochum, Germany}
\author[0000-0002-1748-7367]{Julia Becker Tjus}
\affiliation{Ruhr University Bochum, Faculty for Physics \& Astronomy, Theoretical Physics IV, Plasma-Astroparticle Physics, Bochum, Germany}
\affiliation{Ruhr Astroparticle and Plasma Physics Center (RAPP Center), Ruhr University Bochum, Germany}
\affiliation{Department of Space, Earth and Environment, Chalmers University of Technology, SE-412 96 Gothenburg, Sweden}
\author[0000-0002-9151-5127]{Horst Fichtner}
\affiliation{Ruhr University Bochum, Faculty for Physics \& Astronomy, Theoretical Physics IV, Plasma-Astroparticle Physics, Bochum, Germany}
\affiliation{Ruhr Astroparticle and Plasma Physics Center (RAPP Center), Ruhr University Bochum, Germany}
\author[0000-0002-9881-8112]{Elena Amato}
\affiliation{NAF-Osservatorio Astrofisico di Arcetri, Universita degli Studi di Firenze, Universita degli studi di Firenze}

\begin{abstract}
A complete understanding of the cosmic-ray energy spectrum remains a challenge to theory that must be met by comprehensive modeling efforts. One of these is the subject of the present study, namely, an explanation of the recently discovered spectral hardening at $\sim 300$ GeV with self-consistently treated cosmic-ray diffusion, where self-generated waves resulting from the streaming instability impact the diffusion of high-energy particles. We revisit the corresponding model by \citet{blasi2012}, perform an extensive parameter study, and determine an optimal range of parameters that best fit the cosmic-ray data. We conclude that self-consistently treated cosmic-ray transport remains a competitive alternative to explain the spectral hardening of the cosmic-ray energy spectrum at a few hundred GeV. 
\end{abstract}
\section{Introduction}
Understanding the energy spectrum of cosmic rays in all its complexity remains an important task of contemporary astrophysics, because it contains information about the acceleration as well as propagation of cosmic rays within the Galaxy and beyond, and about the physical properties of the media these high-energy particles traverse. For many years the energy spectrum was considered to have a comparatively simple form mainly characterized by three different power laws \citep{Becker-Tjus-Merten-2020}, namely one in the range from $\sim 10$~GeV to $\sim 3\cdot10^6$~GeV  as a result of (diffusive shock) acceleration at Galactic sources and subsequent propagation in the interstellar medium, a steeper one above the so-called knee up to the ankle at $\sim 4\cdot 10^9$~GeV and finally a third, again flatter one beyond the ankle. While the origin of the range between the knee and the ankle is still under debate, the latter part of the spectrum is presumably comprised of extragalactic cosmic rays.
These power-law sections of the spectrum are bordered at ultra-high energies by the rather sharp 'Greisen–Zatsepin–Kuzmin' cut-off \citep{Greisen-1966, Zatsepin-Kuzmin-1966} and at the low-energy end, if observed at Earth, by a gradual flux decrease due to the so-called modulation caused by the heliosphere \citep{Potgieter-1998}. 

Over the years improved observations have revealed additional features in the cosmic-ray energy spectrum \citep[see, e.g.,][]{Recchia-Gabici-2024}. Above the 'original' knee near $3\cdot 10^6$~GeV a 'second knee' near $10^8$~GeV has been identified \citep{Kampert-2013,Abreu-etal-2021} and is now considered as one of several element-specific knees \citep{DeRujula-2019}. Below the knee, measurements with the DAMPE satellite \citep{DAMPE} confirmed a spectral hardening at $\sim 300$~GeV found before (by PAMELA \citep{PAMELA} as well as AMS02 \citep{Aguilar-etal-2015}) and led to the discovery of a softening at $\sim 13.6$~TeV. These measurements were subsequently confirmed with CALET \citep{Adriani-etal-2023}. 

The present study is devoted to the latter feature, i.e.\ the spectral hardening at $\sim 300$ GeV that may be explained as an effect of cosmic-ray sources in the vicinity of the Sun \citep[][and references therein]{Bhadra-etal-2025, Qian-etal-2025} or of the cosmic-ray transport in the interstellar medium \citep[][and references therein]{Liu2018,Silver_2024} or of deviations of the injection spectra of cosmic rays from a simple power law \citep[see, e.g.,][]{Biermann-etal-2010,Pan_2023}. While the explanations by local sources and deviations in the source spectrum address the hardening and softening at the same time, self-consistent models of the cosmic-ray transport often address only the hardening. 
Out of these alternatives, we study the second by revisiting the idea that this spectral hardening is related to the streaming instability of cosmic rays, which was first studied in the context of spectral breaks in the cosmic ray energy spectrum by \citet{blasi2012}. In difference to test particle propagation models based on the assumption of a spatially dependent diffusion coefficient \citep{Liu2018} with a possibly non-separable energy dependence \citep{Tomassetti2012} or on the assumption of a spatially constant diffusion in combination with re-acceleration and convection \citep{Silver_2024}, \citet{blasi2012} proposed a self-consistent approach taking into account the interaction of the cosmic rays with self-generated turbulence. 

The paper is organized as follows. After we describe and justify the model equations in section~\ref{sec:equations}, we present a comprehensive parameter study in section~\ref{sec:paramstud}. This is followed in section~\ref{sec:MCMC} by a determination of a parameter combination best-fitting the cosmic ray data. In the final section~\ref{sec:disccusion} all findings are critically discussed and corresponding conclusions are drawn. 

\section{Model Equations} \label{sec:equations}
Following the work of \citet{blasi2012} and \citet{Aloisio}, we consider cosmic-ray particles, accelerated by supernova remnants (SNR) and traveling along the field lines of a large-scale magnetic field $\mathbf{B_0}= B_0\mathbf{z}$.
Alongside the regular field, the surrounding plasma exhibits a weakly turbulent component, $\delta B$, in the form of Alfvén waves. Via electromagnetic interaction these waves can serve as scattering centers for the cosmic-ray particles and can be enhanced by the particles in this process. As the wave enhancement depends on the particle distribution, which itself is affected by the Alfvén waves, this so-called streaming-instability can be treated with a coupled system of differential equations, which we introduce in the following.

As a first step, we adapt a one-dimensional flux-tube approximation, where purely parallel diffusion is considered under the assumption that processes influencing the perpendicular direction leave the parallel part unaffected.
The steady-state transport equation describing the cosmic-ray particles then reads
\begin{equation}
    -\frac{\partial}{\partial z}\left[D \frac{\partial f}{\partial z}\right]+v_{\mathrm{A}} \frac{\partial f}{\partial z}-\frac{d v_{\mathrm{A}}}{d z} \frac{p}{3} \frac{\partial f}{\partial p}=q_{\mathrm{CR}}(p,z) = q_0(p)\delta(z)\,,\label{eq:CR_transport}
\end{equation}
where the terms denote spatial diffusion with a momentum-dependent diffusion coefficient $D = D(p)$, advection with the Alfvén speed $v_A= B_0/\sqrt{4\pi n_im_i}$, adiabatic expansion/compression, and a source, respectively.
The isotropic phase-space distribution function $f(p,z)$ is normalized to the total number of particles in the process, $N=\int \dd p \,\dd V\, 4\pi p^2f(p,z) $, where $\dd V$ denotes a volume element.

Secondly, the shock-accelerated particles are assumed to be injected with a power-law spectrum, homogeneously distributed over the Galactic disk, i.e.\ at $z=0$ for the disk structure oriented in (x,y),
\begin{equation*}
    q_0(p)=A \left(\frac{p}{mc}\right)^{-\alpha}\,.
\end{equation*} 
The normalization constant $A$ is determined by the part of the energy of the SN explosion that is transferred to the cosmic rays. A fraction $\xi_{\mathrm{CR}}$ of the total kinetic energy released by the SN, $E_{\mathrm{SN}}$, bears upon the acceleration of particles, multiplication with the SN occurrence rate $\mathcal{R}_{\mathrm{SN}}$ yields the temporally averaged abundance, and division by the area of the Galactic disk, $\pi R_{\mathrm{d}}^2$, accounts for a spatial average. Let 
\begin{equation}
    \epsilon(p)= (\sqrt{(p/(mc))^2+1}-1)mc^2
\end{equation} 
denote the momentum-dependence of the cosmic-ray kinetic energy. Here, we will only consider protons, as the cosmic-ray spectrum in the considered energy range from GeV to TeV is not only dominated by protons, but the proton spectrum can even be separated observationally from the other species so that a comparison of the modeled with the measured proton spectrum is possible. We leave it to future work to add heavier elements, as other effects like a different injecting population need to be considered, see, e.g.\ \cite{biermann2010}. Then, the normalization reads
\begin{align}
    \int q_0(p)\epsilon(p)\, \dd^3 p &= \nonumber \\
    A\int_0^\infty \left(\frac{p}{mc}\right)&^{-\alpha} \epsilon(p)4\pi p^2\, \dd p = \frac{\xi_{\mathrm{CR}}E_{\mathrm{SN}}\mathcal{R}_{\mathrm{SN}}}{\pi R_{\mathrm{d}}^2}\,.
    \label{eq:normalization_A}
\end{align}
After identifying 
\begin{align}
    \mathcal{I}(\alpha)&= \int_0^\infty \left(\frac{p}{mc}\right)^{-\alpha} 4\pi p^2 \left[\sqrt{\left(\frac{p}{mc}\right)^2 +1}-1\right] mc^2 \,\mathrm{d} p \nonumber\\
    &= 4\pi \int_0^\infty x^{2-\alpha}\left(\sqrt{x^2+1}-1\right)\,\mathrm{d} x\,,
\end{align}
where in the last step we substituted $x=p/(mc)$, one finds for the normalization constant 
\begin{align}
    A&= \frac{\xi_{\mathrm{CR}}E_{\mathrm{SN}}\mathcal{R}_{\mathrm{SN}}}{\pi R_{\mathrm{d}}^2 \mathcal{I}(\alpha)c(mc)^4}\nonumber\\ 
    &= \,4\cdot10^{36}\mathrm{cm}^{-5}\mathrm{g}^{-3}\mathrm{s}^2\left(\frac{\xi_{\mathrm{CR}}}{0.1}\right)\left(\frac{E_{\mathrm{SN}}}{10^{51}\,\mathrm{erg}}\right)\nonumber\\&\qquad\times\left(\frac{\mathcal{R}_{\mathrm{SN}}}{(30\,\mathrm{yr})^{-1}}\right) \left(\frac{R_{\mathrm{d}}}{10\,\mathrm{kpc}}\right)^{-2} \left(\frac{\mathcal{I}(\alpha)}{\mathcal{I}(4.3)}\right)^{-1} \,.
\end{align}
Note that the integral in $\mathcal{I}(\alpha)$ diverges for $\alpha\leq4$.

Analogously to the original reference, we further assume that the spatial diffusion coefficient $D(p)$ as well as the Alfvén wave speed $v_A$ are independent of the $z-$coordinate and only demand that the Alfvén waves propagate away from the disk at $z=0$ both below and above, which corresponds to $\dd v_A /\dd z = 2v_A\delta(z)$.

The diffusion coefficient is determined by the magnetic field perturbations $\delta B$ propagating as transversal Alfvén waves, which can resonantly interact with cosmic-ray particles. Any particle that by coincidence traverses a distance of an Alfvén wavelength during one gyration experiences a constant Lorentz force directed parallel to the $\mathbf{B}_0$-field and, as a consequence, is scattered. The resonance condition, therefore, is $k=1/r_{\mathrm{L}}= qB_0/(pc)$, where $k$ denotes the wave number, $r_{\mathrm{L}}$ the Larmor radius of a particle and $q$ its charge.
Assuming, thirdly, the turbulence level to be low, $\delta B \ll B_0$, the parallel diffusion coefficient can be written as
\begin{equation}
    D(p)= \frac{1}{3}r_{\mathrm{L}}(p)v(p)\frac{1}{kW(k)}\,,
    \label{eq:D(p)}
\end{equation}
with the Bohm diffusion coefficient $D_{\mathrm{B}}=r_{\mathrm{L}}v/3$ as the limit of most efficient scattering \citep{SHALCHI2009}. 
The spectral power $W(k)$ is normalized by the turbulence level $\eta$ such that
\begin{equation}
    \int_{k_0}^\infty \dd k\, W(k) =\eta = \frac{\delta B^2}{B_0^2} \,.
    \label{eq:norm_W}
\end{equation}
Not only do the waves have an effect on the cosmic rays and vice-versa, but, likewise, wave-wave interactions can occur, not connected to cosmic rays. As two counter-moving fluctuations encounter each other, they will tend to break up into smaller 'eddies' due to nonlinear couplings. \citet{leith} and  \citet{zhou1990} suggest that the net effect of all these interactions can be considered to be a local energy transfer in $k-$space. We, therefore, describe as a fourth step the evolution of the wave spectrum via a diffusion ansatz:
\begin{equation}
    -\frac{\partial}{\partial k}\left[D_{k k} \frac{\partial W}{\partial k}\right]-\Gamma_{\mathrm{CR}} W=q_{W}(k)\,,\label{eq:wave_transport}
\end{equation}
where we take into account diffusion of the spectral energy density, wave growth and an injecting source.\\
The wave diffusion coefficient is determined by $D_{kk} = k^2/\tau_s$, where the spectral energy transfer time is related to the triple correlation time $\tau_3$ and the nonlinear or eddy turnover time $\tau_{nl}$ through $\tau_3\tau_s = c_K^{-1}\tau_{nl}^2$. We refer to the constant $c_K$
as the \textit{Kolmogorov} constant. 

Like \citet{miller1995}, we consider here the \textit{Kolmogorov} and \textit{Kraichnan} phenomenology: In the former case, the triple correlation time $\tau_3$ is assumed to be of the order of the eddy-turnover time, $\tau_{3,\mathrm{Kolm}}=\tau_{nl}= \lambda/\delta v$, whereas \textit{Kraichnan} allows for the magnetic field to take part, yielding $\tau_{3,\mathrm{Kraich}}=\lambda/v_A$. Taking into account that the velocity fluctuations relate to the Alfvén speed as do the magnetic field fluctuations to the average magnetic field, $\delta v^2/v_A^2=\delta B^2/B_0^2$, and that $\lambda\approx 1/k$, we can, analogously to \citet{Aloisio}, write 
\begin{equation}
    D_{kk}= c_Kv_Ak^{\alpha_1}W(k)^{\alpha_2}\,,
    \label{eq:D_kk}
\end{equation} 
with $(\alpha_1, \alpha_2)=(7/2,1/2)$ in case of \textit{Kolmogorov} phenomenology and $(4,1)$ for \textit{Kraichnan}. The first term in Eq.(\ref{eq:wave_transport}) consequently describes wave cascading.

Wave growth, in our model, occurs due to the above-mentioned streaming instability, as particles resonantly interact with and get scattered by Alfvén waves. For example, \citet{Kulsrud1969} state that the interaction is, on a larger scale, associated with a momentum transfer. Given an isotropic distribution, the effect of momentum loss and gain by the particles would balance. However, in the presence of a gradient, the streaming instability can occur, leading to wave growth. The associated growth rate for our purpose can be obtained by pitch-angle averaging the expression given by \citet{skilling1971},
\begin{equation}
    \Gamma_{\mathrm{CR}}(k) = \frac{16\pi^2}{3}\frac{v_A}{kW(k)B_0^2 }\left[p^4v(p)\frac{\partial f}{\partial z}\right]_{p=qB_0/kc}\,, \label{eq:Gamma_CR}
\end{equation}
where $v(p)$ is the speed of a particle with momentum $p$ and $\partial f/\partial z$ the gradient in $z$-direction.
The source term of Eq.~\eqref{eq:wave_transport} in our setup is considered to be $q_W(k)\propto \eta\delta(k-k_0)$, injecting turbulence only on an outer scale $k_0^{-1}$.

The system of equations (\ref{eq:CR_transport}) and (\ref{eq:wave_transport}) is nonlinear for the diffusion coefficient in the particles' transport equation being dependent on the wave spectrum through Eq.\,(\ref{eq:D(p)}) and the wave spectrum in turn being determined by the particle distribution function through the wave growth, Eq.\,(\ref{eq:Gamma_CR}).\\
Owing to the simplifying assumptions, we can find an analytical solution to the particles' transport equation. Imposing the boundary conditions $f(z=\pm H)=0$, where $H$ denotes the Galactic height, it reads
\begin{equation}
    f(z, p)=f_0(p) \frac{1-e^{-\zeta(1-|z| / H)}}{1-e^{-\zeta}}\,, \quad \zeta(p) \equiv \frac{v_{\mathrm{A}} H}{D(p)}\,.
    \label{eq:f(z,p)}
\end{equation}
The momentum-dependent part $f_0(p)$ can be found by integrating Eq.(\ref{eq:CR_transport}) in the range $z=(0^- - 0^+)$,
\begin{equation}
    -2D(p)\left[\frac{\partial f}{\partial z}\right]_{z=0^+} -\frac{2}{3}v_Ap\frac{\dd f_0}{\dd p}= q_0(p)\,,
    \label{eq:dfdz}
\end{equation}
from which in turn, an implicit solution can be derived as
\begin{equation}
        f_0(p) = \frac{3}{2v_A}\negthickspace\int^\infty_p\frac{\dd p''\,q_0(p'')}{p''}\exp\negthickspace\left[\int^{p''}_p\negthickspace\frac{3\,\dd p'}{p'\left(1-\mathrm{exp}\left[\zeta(p)\right]\right)}\right]\,.
        \label{eq:f_int}
\end{equation}
For this, we used that 
\begin{equation}
    \left[\frac{\partial f}{\partial z}\right]_{z=0^+}= \frac{v_Af_0(p)}{D(p)}\frac{1}{1-\exp\left[\zeta(p)\right]}\,,
    \label{eq:dfdz2}
\end{equation}
obtained by spatially differentiating Eq.\,(\ref{eq:f(z,p)}), and with this derivative inserted, integrating Eq.\,(\ref{eq:dfdz}).\\
Likewise, an implicit solution to the wave equation (\ref{eq:wave_transport}) can be found after integrating twice,
\begin{align}
    W(k) = \Bigg[ \negmedspace\left(-\frac{1+\alpha_2}{c_K v_A}\int_k^\infty\negmedspace\frac{\mathrm{d} k'}{k'^{\alpha_1}}\negmedspace\int_{k_0}^{k'}\negthickspace \mathrm{d} k''\,\Gamma_{\mathrm{CR}}(k'')W(k'')\negmedspace \right)  \nonumber\\
    + \left(c_3 k^{1-\alpha_1}\right)\negmedspace\Bigg]^{\frac{1}{1+\alpha_2}}\,.
    \label{eq:w_int}
\end{align}
The integration constant $c_3$ here is determined by the requirement that in the absence of resonating cosmic-ray particles the expected spectra, namely \textit{Kolmogorov}- or \textit{Kraichnan}-like, are found. Then, we find $c_3=W_0^{1+\alpha_2}k_0^{\alpha_1-1}$ with $W_0=(s-1)\,k_0^{-1}\eta$ and the index
$s= (\alpha_1-1)/(\alpha_2+1)$.\\
We solve the coupled differential equations (\ref{eq:CR_transport}) and (\ref{eq:wave_transport}) by numerically integrating the implicit solutions (\ref{eq:f_int}) and (\ref{eq:w_int}) with a trapezoidal method.
In concrete terms, we divide each integral into $2\times 10^6$ narrow intervals. Each of the inner integrals are solved with the {\fontfamily{pcr}\selectfont SciPy} \citep{scipy} built-in \texttt{cumulative trapezoid} method, whereas the outer integrals are solved with the {\fontfamily{pcr}\selectfont SciPy} \texttt{trapezoid} methods for momentum $p$ and wavenumber $k$ in the ranges $[10^{-2},10^7]\,$GeV/c and $[k_0,10^{-10}]\,\mathrm{cm}^{-1}$, respectively. In order to minimize the runtime, the implicit solutions are evaluated at 300 points, which we note is less than the number of integrands, and interpolated in between. As starting conditions, we take the distribution function to be $f_{i=0}(p)=q_0(p)H/(2D(p))$ and assume a wave spectrum according to $W_{i=0}=W_0 \left(k/k_0\right)^{-s}$.
The integrations are repeated in such a way that $f_i$ is calculated according to Eq.\,(\ref{eq:f_int}) with $D(p)$ taken from the previous iteration, $f_i(D_{i-1})$. Hereafter, $W_i$ is computed as in Eq.\,(\ref{eq:w_int}), where $\partial f/\partial z$ is derived via Eq.\,(\ref{eq:dfdz2}) with $f(p)$  taken from the same iteration, $W_i(f_i)$. This iterative procedure is repeated until convergence is reached, which for our purposes is given when the maximal relative deviation of two consecutive iterations, $\left\lVert f_i/f_{i-1}\right\rVert_\mathrm{max}$ and $\left\lVert W_i/W_{i-1} \right\rVert_\mathrm{max}$, amounts to less than 0.5\%.

\section{Quantitative Parameter study}\label{sec:paramstud}
Having introduced the model, there are different dependencies that are useful to investigate prior to performing a systematic Markov Chain Monte Carlo (MCMC) fit. We fix the values for the Galactic height $H$ and the turbulence injection scale $k_0^{-1}$ to typical values of $H=4\,$kpc and $k_0^{-1}=50\,$pc. As already discussed by \citet{blasi2012}, a variation of $k_0^{-1}$ does not affect the position of the spectral break. The remaining parameters are the large-scale magnetic field strength $B_0$, its turbulence level $\eta$, the Alfvén speed $v_A$, 
the cosmic-ray injection slope $\alpha$ and factor $A$, which itself is composed of multiple quantities, the \textit{Kolmogorov constant} $c_K$.

\begin{figure}
    \centering    \includegraphics[width=\linewidth]{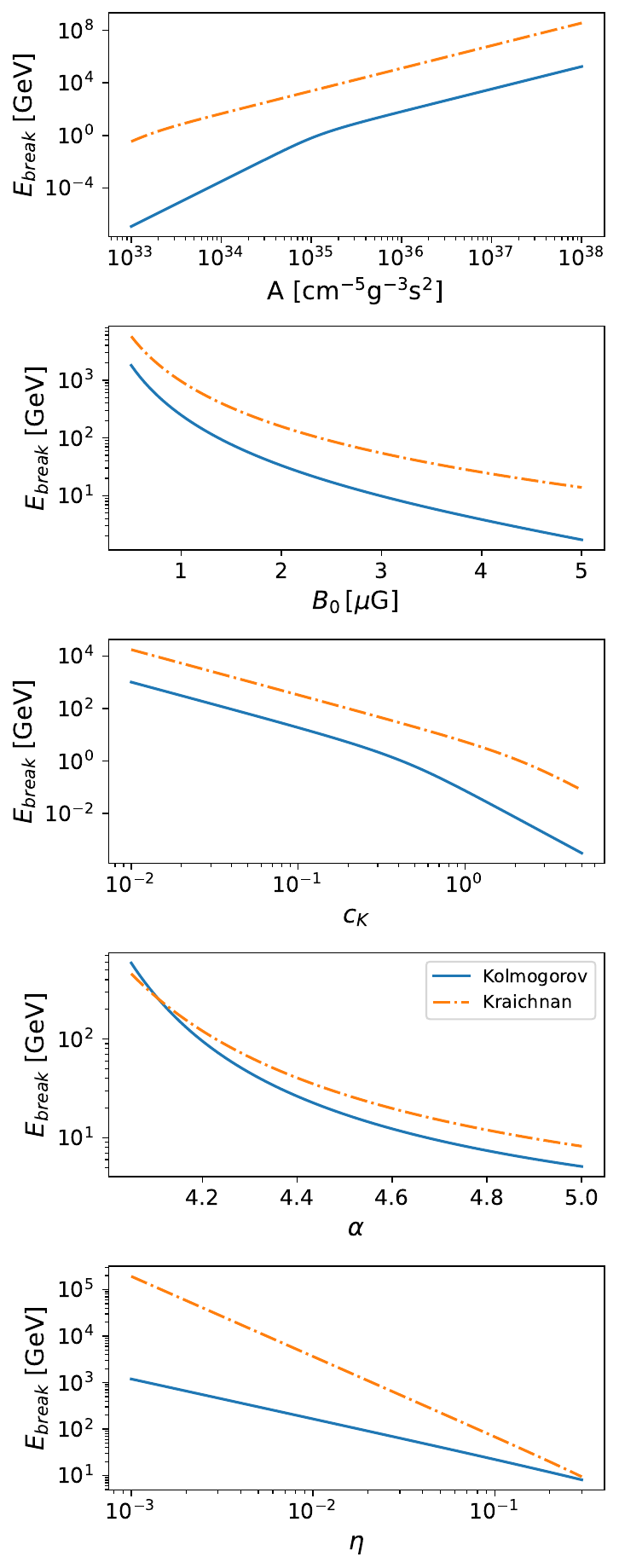}
    \caption{Comparison of the expected break energies for different values of $A,\,\eta,\,c_K,\,\alpha,$ and $B_0$ in case of either \textit{Kolmogorov}- or \textit{Kraichnan}-type of wave cascade. The respective fixed values are chosen as in Fig.\,\ref{fig:default_params}. 
    }
    \label{fig:break_pos}
\end{figure}

In order to roughly estimate the expected position of the spectral break caused by the streaming instability, we
use that the wave growth rate $\Gamma_{\mathrm{CR}}$, given by Eq.\,(\ref{eq:Gamma_CR}) needs to prevail over the counteracting wave cascading with $\Gamma_{\mathrm{D}}\approx D_{kk}/k^2$.
Approximating $\partial f/\partial z\approx -f_0/H$ with the distribution function chosen to be the solution to the diffusion-dominated transport equation at $z=0$, $f_0=q_0H/(2D)$, and where the diffusion coefficient according to Eq.\,(\ref{eq:D(p)}) is determined by the external wave spectrum $W=(s-1)k_0^{s-1}\eta k^{-s}$, the momentum at which a transition from wave cascading to growth occurs can be calculated as
\begin{align}
    p_{\mathrm{break}} = \Bigg\{ \frac{8\pi^2}{B_0^2c_K}&k_0^{\alpha_2(1-s)}\left[\eta\left(s-1\right)\right]^{-\alpha_2}\nonumber\\
    \times \,A&(mc)^\alpha\left(\frac{qB_0}{c}\right)^{3-\alpha_1+s\alpha_2}
    \Bigg\}^{\frac{1}{\alpha+s\alpha_2-\alpha_1-1}}\,.
    \label{eq:p_break}
\end{align}
We display the different dependencies of the break energies on various parameters in Fig.\,\ref{fig:break_pos}.
In almost all cases the break energies are higher with the \textit{Kraichnan}-like cascade. An increase of the injection factor $A$ leads to increasing break energies. The opposite is true for the remaining parameters. The visible changes of slope are due to the transition from the relativistic to the non-relativistic regime.

\begin{figure}
    \centering
    \includegraphics[width=1.0\linewidth]{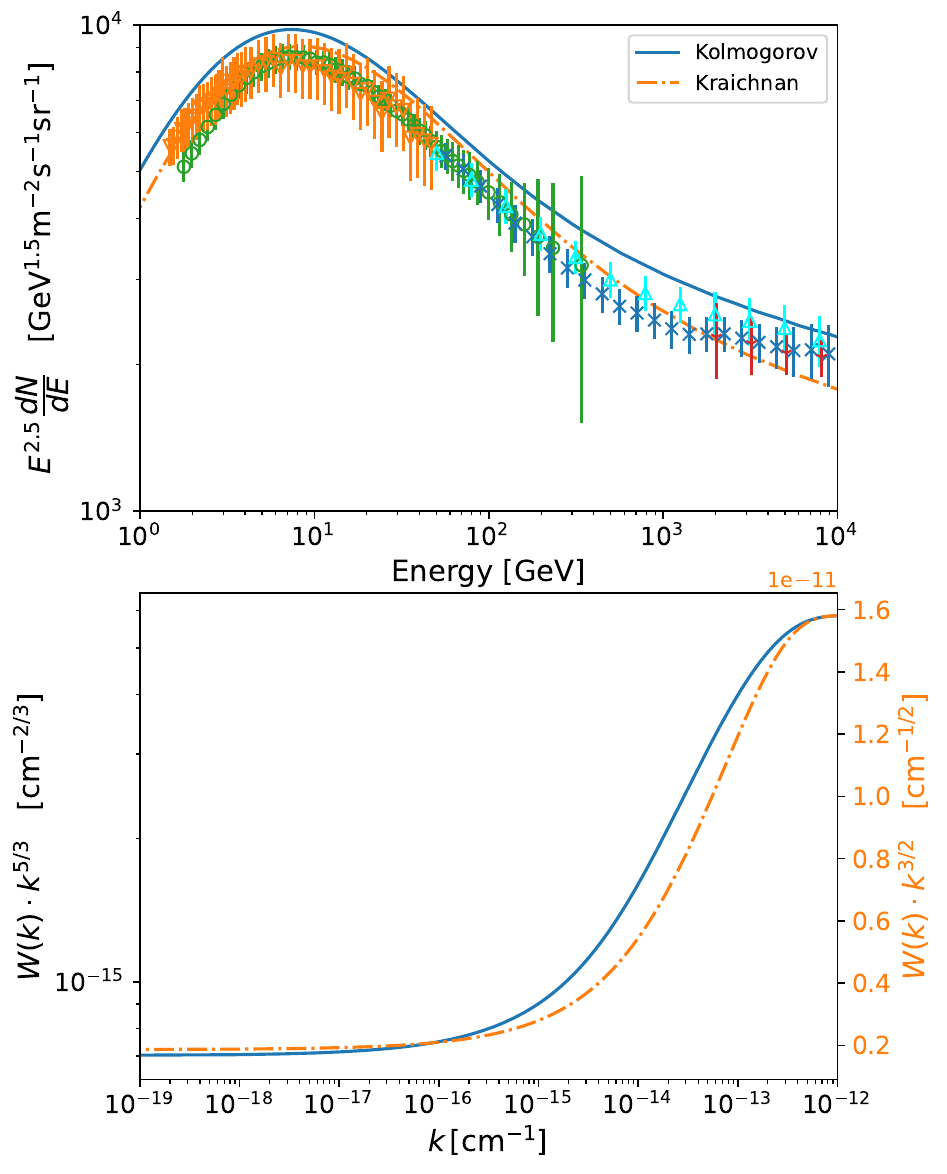}
    \caption{Comparison of the cosmic-ray fluxes (top) and wave spectra (bottom) obtained with the default parameter sets. In the \textit{Kolmogorov} case,  we choose $A=10^{36}\,\mathrm{cm}^{-5}\mathrm{g}^{-3}\mathrm{s}^2$, $\alpha=4.25$, $\eta=0.03$, $v_A=25\,\mathrm{km\, s}^{-1}$, $B_0=1.6\,\mu\mathrm{G}$, $
     H= 4\,\mathrm{kpc}, k_0^{-1}=50\,\mathrm{pc}$,  whereas for \textit{Kraichnan} it is $A=10^{34.5}\,\mathrm{cm}^{-5}\mathrm{g}^{-3}s^2$, $\alpha=4.08$, $\eta=0.04$, $v_A=1\,\mathrm{km\,s}^{-1}$, $B_0=1.5\,\mu$G, $H=4\,$kpc, $k_0^{-1}=50\,$pc.}
    \label{fig:default_params}
\end{figure}

In the following, we perform a quantitative parameter study to obtain a first idea of the various parameters' influences on the particle distribution function and the wave spectrum. The default parameter set is chosen such that the models resemble observational cosmic-ray data (Figure \ref{fig:default_params}).
From there, we successively vary one parameter at a time.
Figures \ref{fig:param_kolm} and \ref{fig:param_kraich} show the results for a \textit{Kolmogorov} and a \textit{Kraichnan} phenomenology, respectively. Figure \ref{fig:param_kolm} shows the behavior of the cosmic-ray energy spectrum (left) and magnetic wave spectrum (left) at the variation of the different parameters. The estimated break energy is indicated in Fig.\,\ref{fig:param_kolm} and \ref{fig:param_kraich} by vertical lines. The following trends are the same for figure \ref{fig:param_kraich}: 
\begin{figure*}
    \centering
    \includegraphics[width=\textwidth, height=0.95\textheight, keepaspectratio]{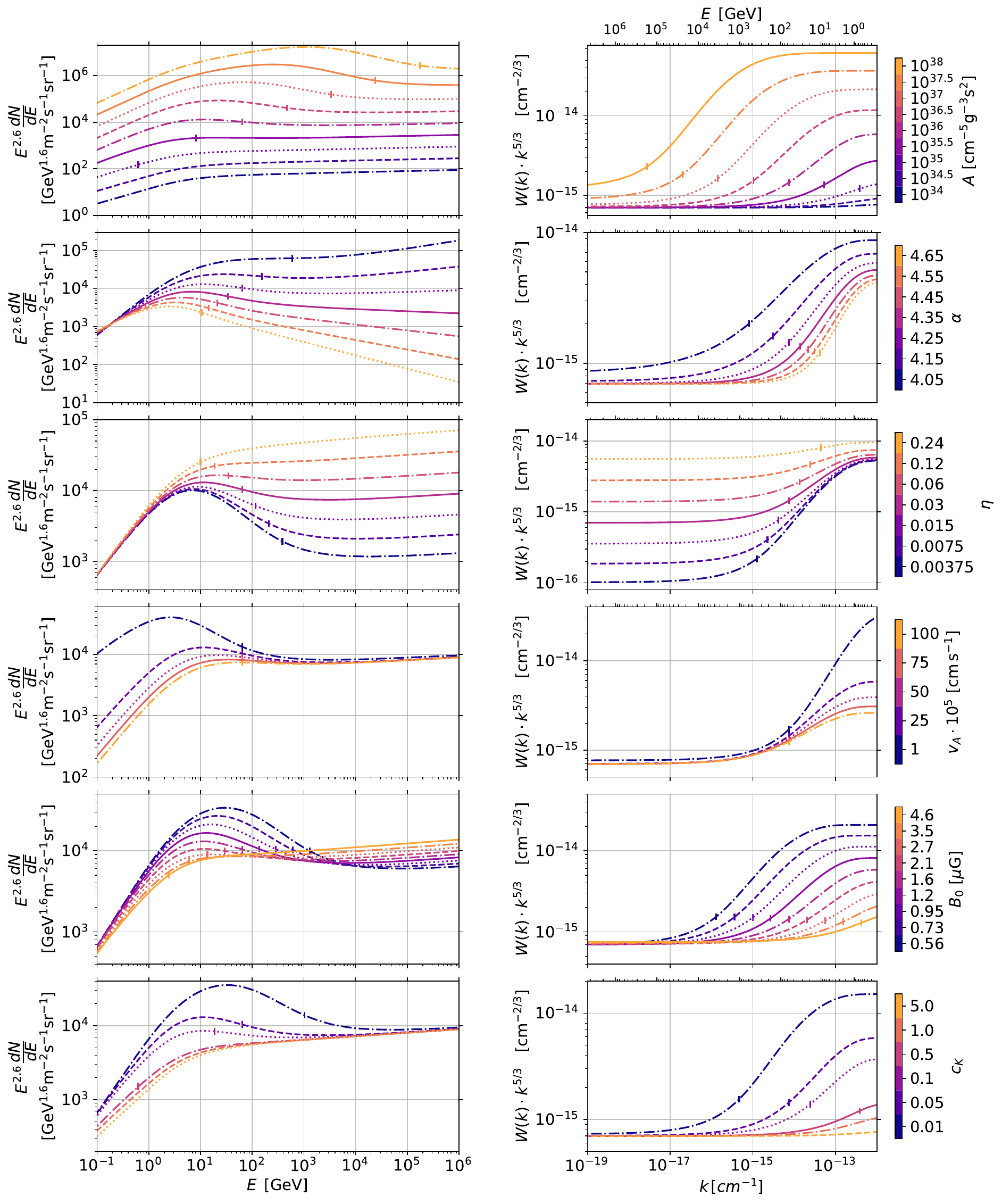}
    \caption{\textit{Kolmogorov}. 
    We vary the model parameters around the default parameter set (as specified in Fig.\,\ref{fig:default_params}), one in each row while keeping the rest fixed. Line styles are changing from \textit{dashed-dotted}, \textit{dashed}, \textit{dotted}, \textit{solid} with increasing parameter value. The energies for the right plots are derived as $E=(\sqrt{1+(qB_0/(kmc^2))^2}-1)mc^2$, except for the second last row, where we varied $B_0$. The expected breaks according to Eq.\,(\ref{eq:p_break}) are denoted by small vertical lines.}   \label{fig:param_kolm}
\end{figure*}
\begin{figure*}
    \includegraphics[width=\textwidth, height=0.95\textheight, keepaspectratio]{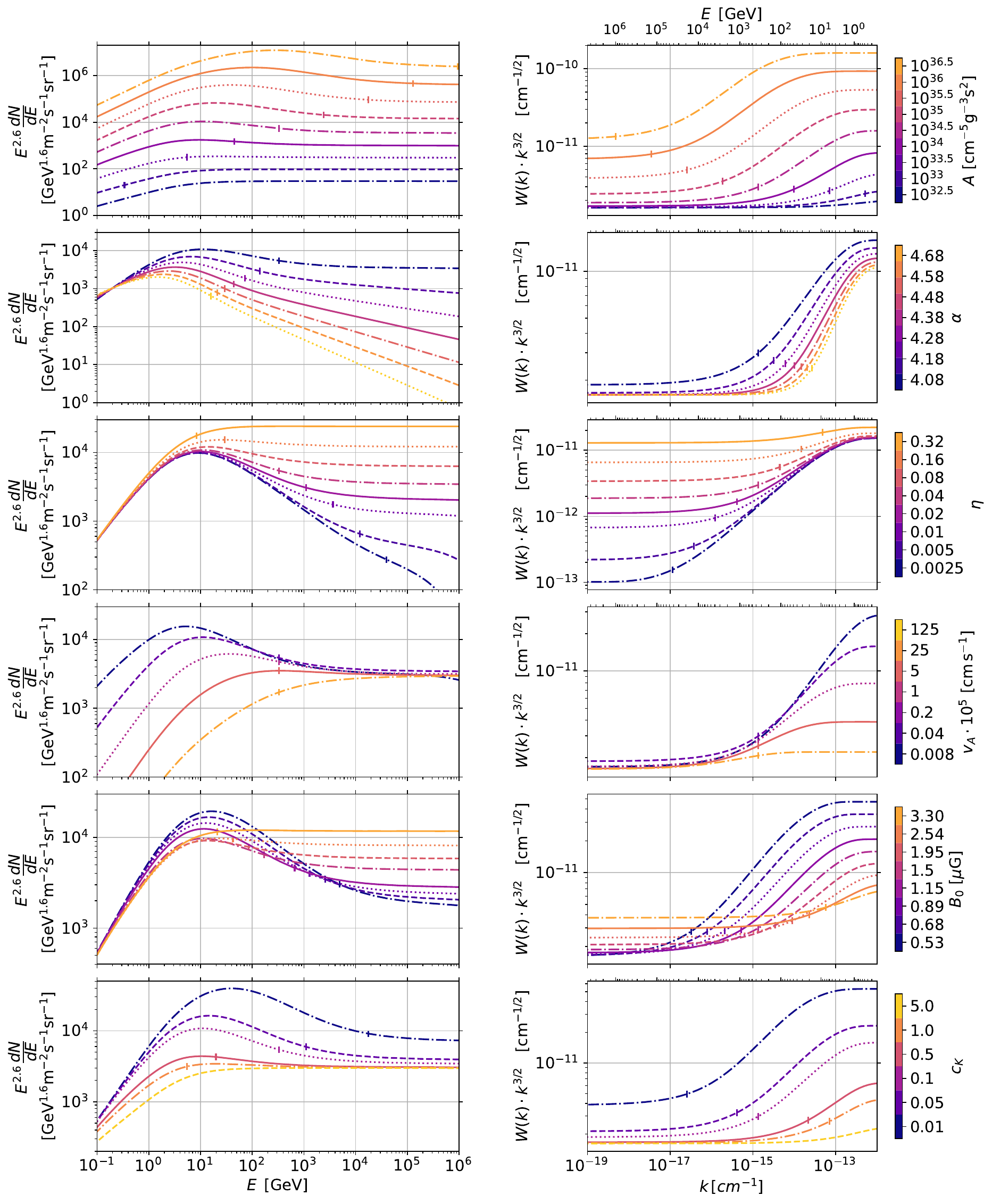}
    \caption{\textit{Kraichnan}. Starting from the default values, the parameters are changed in a similar manner as in Fig.\,\ref{fig:param_kolm}, with the line styles kept in the same changing order. Again, small vertical lines denote the expected breaks due to the streaming instability.}    \label{fig:param_kraich}     
\end{figure*}

\begin{itemize}
\item[-] \textbf{Normalization of the cosmic-ray injection spectrum $A$:} Given the wide range of the normalization constant $A$, we see that apart from shifting the particle spectrum by a constant factor of $10^{0.5}$ -- since the absolute amount of particles in the system is being changed -- also the energy range in which streaming-instability occurs is shifted. The higher $A$, the higher are also the energies at which wave growth becomes the predominant effect. This is visible both in the spectral break of the particle distribution and even more clearly in the wave spectrum. 
\item[-] \textbf{Variation of $\alpha$:} The variation of the spectral index $\alpha$ mainly affects the steepness of the particle distribution function while all functions almost meet in one point at about $mc\left(\alpha_i/\alpha_j\right)^{1/(\alpha_j-\alpha_i)}\approx 0.26$ GeV, taking the highest index, $\alpha_j$ and lowest, $\alpha_i$, under consideration in the advection-dominated solution. A lower value of $\alpha$ corresponds to a flatter distribution $f$ (and thus $\dd N/\dd E$) and more particles that are available to resonantly interact with the abundant magnetic field perturbations. With decreasing spectral index, growth sets in at higher energies. In addition, the level of wave power at the high-energy end of the streaming instability regime remains higher. For the two lowest values of $\alpha$, an even higher turbulence level in the high-energy range is obtained.
\item[-] \textbf{Turbulence level $\eta$:} The initial turbulence level mainly determines the high-energy regimes. At lower initial wave power, the streaming instability prevails sooner, as expected according to Eq.\,(\ref{eq:p_break}). Independent of $\eta$, the final wave power approaches almost the same level at higher wave numbers and lower energies, respectively. On the particle side, this translates into a larger particle population at higher energies due to the dependency of the diffusion coefficient (\ref{eq:D(p)}) on $W(k)$, the higher $\eta$, or $\delta B$ respectively, and the merging of the distribution functions toward lower energies.
\item [-] \textbf{Influence of Alfvén speed $v_A$:} In comparison, the variation of the Alfvén speed $v_A$, which is at fixed magnetic field $B_0$ associated with the variation of the background ion density $n_i$, has a somewhat reverse effect. While the high-energy regime is unaltered, at lower energies the spectra spread apart. From Eq.(\ref{eq:dfdz}), upon neglecting the diffusion term -- since at lower energies advection dominates -- the solution $f_0=3A/(2v_A\alpha)\,(p/(mc))^{-\alpha}$ can be found. Consistently, a lower speed $v_A$, which corresponds to a higher density $n_i$, leads to a higher cosmic-ray density. The particles lose their energy until they advect with Alfvén speed. The position of the expected spectral break energy remains the same for all alterations since the latter is independent of $v_A$.
\item[-] \textbf{Background magnetic field $B_0$:} The ambient magnetic field strength $B_0$ was, starting from the default value $1.6\,\mu$G, successively increased/decreased by a factor of $1.3$, leading to an almost equidistant shift of wave spectra at lower energies. Increasing $B_0$ in our model affects only the growth rate $\Gamma_{\mathrm{CR}}\propto B_0^{-2}$, leading to more wave excitation for smaller values of $B_0$. 
Due to the normalization of the wave spectrum to a fixed amount of turbulence, the smaller wave growth at higher $B_0$ is necessarily associated with a higher wave abundance at higher energies. This effect is much more pronounced in the \textit{Kraichnan} case (Fig. \ref{fig:param_kraich}).
\item[-] \textbf{Kolmogorov constant $c_K$:} This constant makes wave diffusion less effective the smaller its value is. As a consequence, excited waves tend to remain more abundant.
\end{itemize}
The plots further indicate that the estimated break energies seem to be rather underestimated. 
%

%
\section{Fitting the Galactic Environment parameters} \label{sec:MCMC}
\subsection{Method}
In this section, we use a quantitative fitting procedure to find the parameter combination that is best suited to fit currently available observational cosmic-ray data. The use of a Markov Chain Monte Carlo (MCMC) algorithm \citep{mcmc} additionally allows us to determine a range in which a parameter combination still provides a sufficiently good fit, as well as correlations between particular parameters.
As suggested by \cite{likelihood}, we assume that the deviation of the values obtained by our model, $y_{i,\mathrm{model}}(d_j)$, which is constituted by the model parameters $d_j$, from the given data, $y_{i,\mathrm{data}}$, with combined statistical and systematic error, $y_{i,\mathrm{err}}$, is Gaussian-distributed, so that our conditional probability function for the $i$-th datapoint reads 
\begin{align}
    p_i(y_{i,\mathrm{data}}|d_j&,y_{i,\mathrm{err}}) \nonumber \\
    = \frac{1}{\sqrt{2\pi y_{i,\mathrm{err}}^2}}&\exp\left[-\frac{(y_{i,\mathrm{data}}-y_{i,\mathrm{model}}(d_j))^2}{2y_{i,\mathrm{err}}^2}\right]\,. \label{eq:likelihood}
\end{align}
This choice of the maximum-entropy distribution accounts for our lack of specific knowledge about the model's outcome. The likelihood of the parameters, given the independence of each conditional probability function from another, is simply the product of all these functions, $\mathcal{L}=\prod_{i=1}^{N}p_i(y_{i,\mathrm{data}}|d_j,y_{i,\mathrm{err}})$, $N$ being the number of data points under consideration.
For the best model fit, this likelihood should be maximal. Consequently, if we take the logarithm of Eq.\,(\ref{eq:likelihood}), we then need to maximize
\begin{align}
    \ln \mathcal{L}(y_{i,\mathrm{data}}|&d_j,y_{i,\mathrm{err}})\nonumber\\
    = &-\frac{1}{2}\sum_{i=1}^N
    \left[\frac{\left(y_{i,\mathrm{data}}-y_{i,\mathrm{model}(d_j)}\right)^2}{y^2_{i,\mathrm{err}}}\right] \nonumber \\
    &- \frac{1}{2}\sum_{i=1}^N \ln \left(2\pi y_{i,\mathrm{err}}^2\right)\,.
    \label{eq:ln_likelihood}
\end{align}
Given that the number of free parameters in the model is significant, we restrict some of them to reasonable fixed values. 

Since the aim is to explain the spectral break observed at about 300 GeV, we choose the CALET data \citep{CALET} to be our input values, $y_{i,\mathrm{data}}$. Due to small error bars of the AMS-02 data, the fitting of the lower energy part of the cosmic-ray spectrum would have prevailed with a less prioritized fit of the spectral break as a consequence, which is why we spare it here. The spectral index $\alpha$ is set such that it matches the slope of the CALET data above the break in the diffusion-dominated limit, where $f(p)\propto p^{-\alpha}/D(p)$. In case of the \textit{Kolmogorov} phenomenology, this implies $\alpha=4.25$, while \textit{Kraichnan} suggests $\alpha=4.08$. Both choices are in agreement with recent research advocating for cosmic-ray spectra steeper than $p^{-4}$ \citep[e.g.,][]{BELL2013}. 
The injection scale $k_0^{-1}$ of the turbulence is fixed to $50\,$pc, since this parameter appears to mainly affect the total amount of initial wave power and not the spectral break. The halo height is fixed to a typical value of $H=4\,$kpc. \\ 
On the remaining parameters, we impose uniform priors: 
\begin{equation}
    \pi(d_j)=\begin{cases}
        \frac{1}{d_{j,\mathrm{max}}-d_{j,\mathrm{min}}}, & \mathrm{for }\,\, d_{j,\mathrm{min}}<d_j<d_{j,\mathrm{max}}\,,\\
        0, &\mathrm{otherwise}\,.
    \end{cases}
\end{equation}
We determine the ranges of the parameters based on the quantitative parameter study performed in Sect.\,\ref{sec:paramstud}.
The optimal value for the Alfvén speed $v_A$ is assumed to be found in a range $[10^5,10^8]$ cm/s, the turbulence level $\eta\in[10^{-3},10^1]$, the mean magnetic field $B_0\in[0.6,3.0]\,\mu$G, the \textit{Kolmogorov} constant $c_K\in[10^{-2},10^1]$ and the cosmic-ray injection factor $A\in[10^{34}, 10^{37}]\,\mathrm{cm}^{-5}\mathrm{g}^{-3}\mathrm{s}^2$ in the \textit{Kolmogorov} case and $v_A\in [10^3,10^5]$ cm/s, $\eta\in[10^{-3},10^1]$, $B_0\in[0.6,3.0]\,\mu$G, $c_K\in[10^{-2},10^1]$ and $A\in[10^{33.5},10^{35.5}]\,\mathrm{cm}^{-5}\mathrm{g}^{-3}\mathrm{s}^2$ in the \textit{Kraichnan} case.  
The posterior probability distribution according to Bayesian inference is
\begin{align}
    P(y_{i,\mathrm{model}}|&d_j,y_{i,\mathrm{err}})\nonumber\\\propto\,& \mathcal{L}(y_{i,\mathrm{data}}|d_j,y_{i,\mathrm{err}}) \,\pi(d_j)
    \label{eq:post_prob}
\end{align}
which needs to be maximized. With regard to numerical stability, we maximize the logarithm of this expression.
\subsection{Results}
\begin{figure*}
    \centering
    \includegraphics[width=0.65\linewidth]{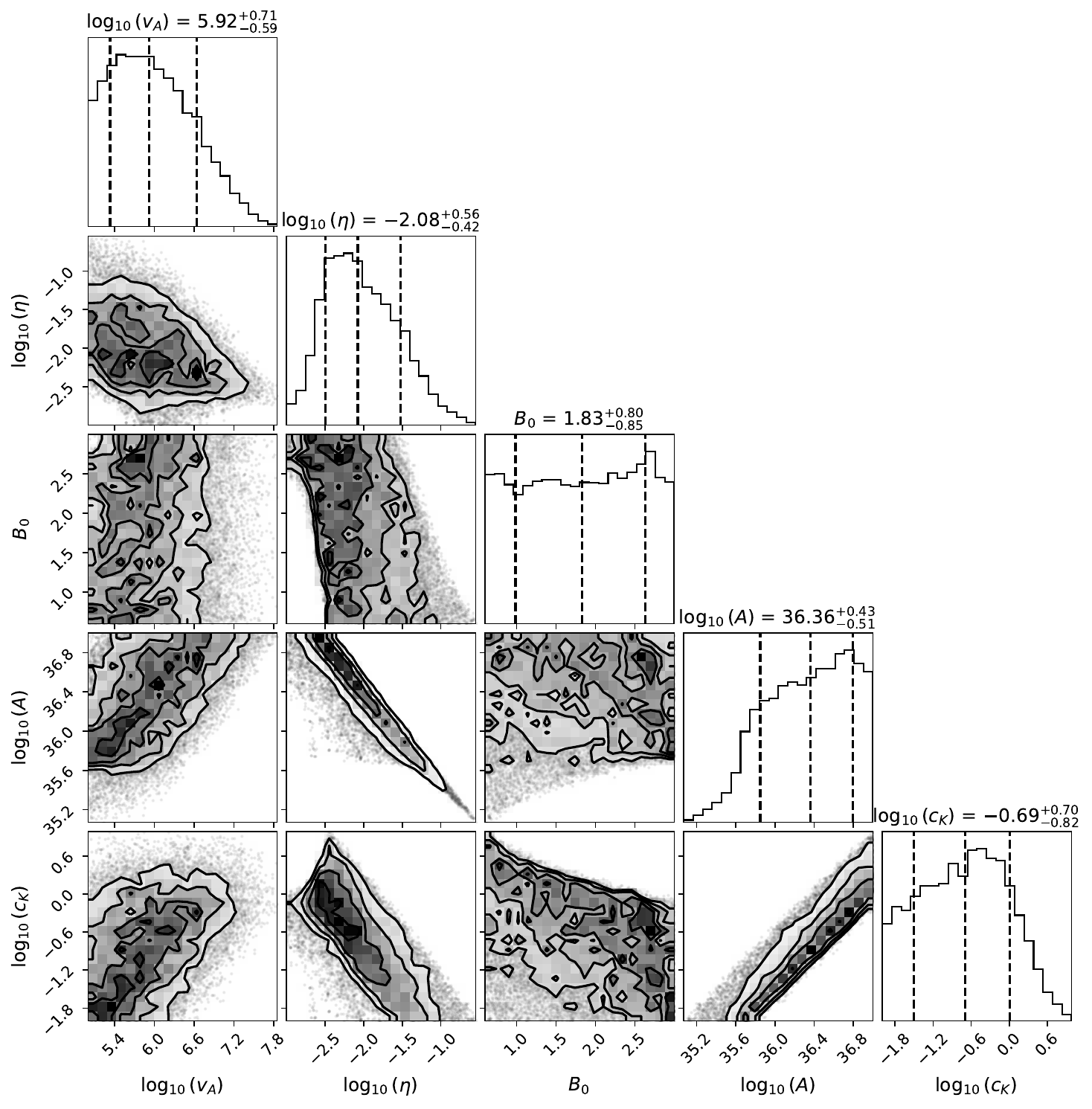}
    \caption{Corner plot of our parameters for a \textit{Kolmogorov}-type cascade. On the diagonal, the marginalized distributions of the Alfvén speed $v_A$, turbulence measure $\eta$, mean magnetic field $B_0$, cosmic-ray normalization constant $A$ and Kolmogorov constant $c_K$ are displayed with the dashed lines denoting the 68\%-credible areas. The remaining panels show the marginalized two-dimensional distributions. }
    \label{fig:corner_kolm}
\end{figure*}
\begin{figure*}
    \centering
    \includegraphics[width=0.65\linewidth]{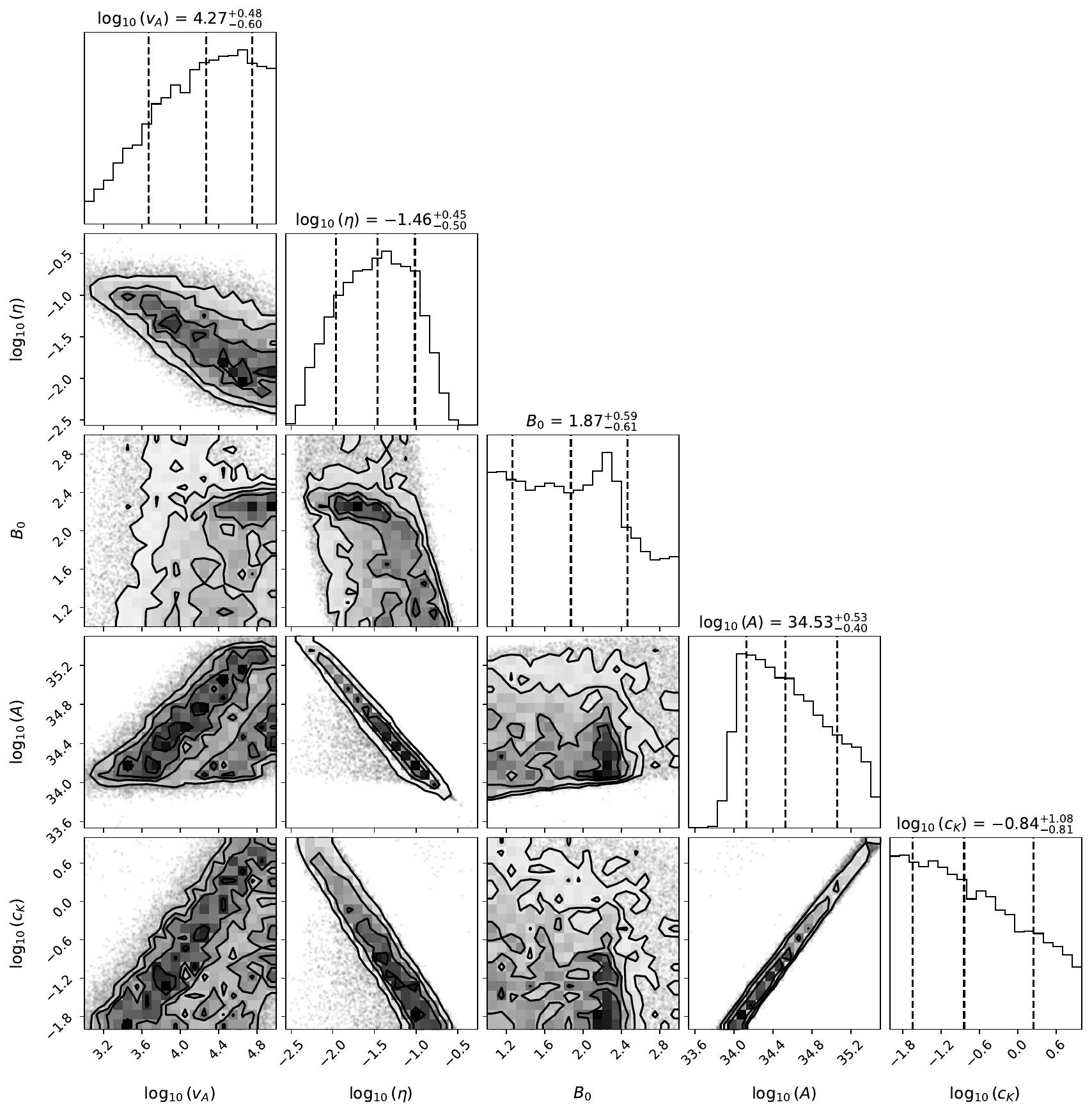}
    \caption{Corner plot for a \textit{Kraichnan}- type cascade. As in Figure \ref{fig:corner_kolm}, the one-dimensional marginalized distributions with the 68\%-credible areas are shown, as well as the marginalized two-dimensional distributions.}
    \label{fig:corner_kraich}
\end{figure*}
Figures \ref{fig:corner_kolm} and
\ref{fig:corner_kraich} show the resulting corner plots, i.e., one- and two-dimensional projections of the posterior probability, the product of our conditional probability function and the priors. For these, we used the {\fontfamily{pcr}\selectfont corner} package \citep{cornerplot} available for {\fontfamily{pcr}\selectfont Python}. 
The solid lines enclosing differently sized and shaded areas denote the 1 -, 2 - and 3 - $\sigma$ intervals.
We find most prominent correlations between the turbulence level $\eta$ and the normalization constant $A$, $\eta$ and the Kolmogorov constant $c_K$, and thus, also between $A$ and $c_K$. Increasing $A$ or $c_K$ is associated with a decrease of $\eta$, which is reasonable since the former two parameter variations lead to greater wave abundance that needs to be balanced by a decrease of pre-existing turbulence. Similarly plausible, $A$ and $c_K$ are correlated in such a way that increasing $A$ requires a decrease of $c_K$.
The ambient magnetic field strength appears to be the least constrained parameter.
Table \ref{tab:median_values} summarizes the median values resulting from the MCMC analysis and the respective $1-\sigma$ intervals.
Due to the various correlations that the parameters exhibit and the fact that most of the 1-d marginalized distributions are not symmetrical, using the median value for each of them in our model does not provide the best fit to the cosmic-ray data. This is rather achieved by choosing the parameter combination that leads to the minimal posterior probability distribution.
\begin{table*}
\begin{center}
\begin{tabular}{ l | c c c c c c c c} 
parameter & symbol & unit & \multicolumn{2}{c}{tested range} & \multicolumn{2}{c}{median value} & \multicolumn{2}{c}{1-$\sigma-$interval}\\
& && \textit{Kolmogorov} & \textit{Kraichnan}& \textit{Kolmogorov} & \textit{Kraichnan} & \textit{Kolmogorov} & \textit{Kraichnan} \\
 \hline
 Alfvén speed & $\log_{10}v_A$ & cm $\mathrm{s}^{-1}$& $[5,8]$ & $[3,5]$ &$5.92$ &$4.27$ & $[5.33,6.63]$ & $[3.67,4.75]$\\ 
 halo height &$H$ &kpc & 4 & 4 & $-$&$-$ & $-$&$-$\\ 
 turbulence level& $\log_{10}\eta$ & $-$& $[-3,1]$ & $[-3,1]$ & $-2.08$ & $-1.46$ & $[-2.50,-1.52]$ & $[-1.96,-1.01]$\\ 
 injection scale &$k_0^{-1}$& pc & 50 & 50 &$-$&$-$&$-$&$-$\\
 background field & $B_0$ & $\mu$G & $[0.6,3.0]$& $[0.6,3.0]$ &1.83 & 1.87 & $[0.98,2.63]$ & $[1.26,2.46]$\\
 injection factor &$\log_{10}A$ & $\mathrm{cm}^{-5}\mathrm{g}^{-3}\mathrm{s}^2$ & $[34,37]$& $[33.5,35.5]$& $36.36$ & $34.53$ & $[35.85,36.79]$ & $[34.13,35.06]$\\
 injection slope &$\alpha$ & $-$ &4.25 & 4.08 &$-$&$-$&$-$&$-$\\
 Kolmogorov constant & $\log_{10}c_K$ & $-$ & $[-2,1]$ & $[-2,1]$ & $-0.69$& $-0.84$ & $[-1.51,0.01]$&$[-1.65,0.24]$ \\
 \hline
\end{tabular}
\caption{Summary of all model parameters, their allowed ranges we used and the median results with their 1-$\sigma-$interval.}
\label{tab:median_values}
\end{center}
\end{table*}
\begin{figure}
    \centering
    \includegraphics[width=\linewidth]{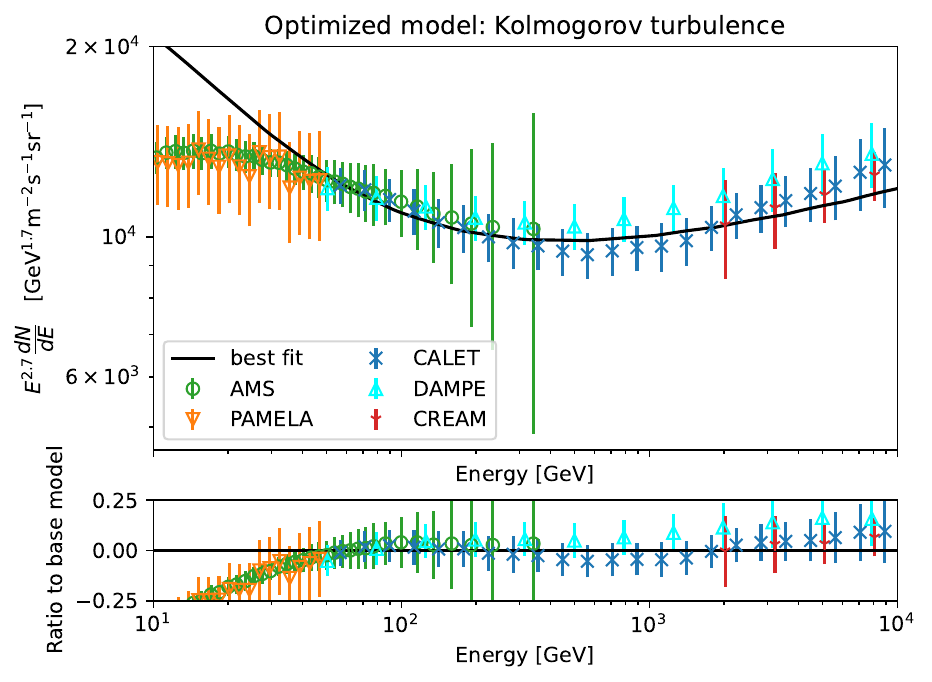}
    \caption{Cosmic-ray particle spectrum with the best-fit parameters for the \textit{Kolmogorov} case compared to data from AMS-02 \citep{AMS}, PAMELA \citep{PAMELA}, CALET \citep{CALET}, DAMPE \citep{DAMPE} and CREAM \citep{CREAM}.}
    \label{fig:f_kolm}
\end{figure}
\begin{figure}
    \centering
    \includegraphics[width=\linewidth]{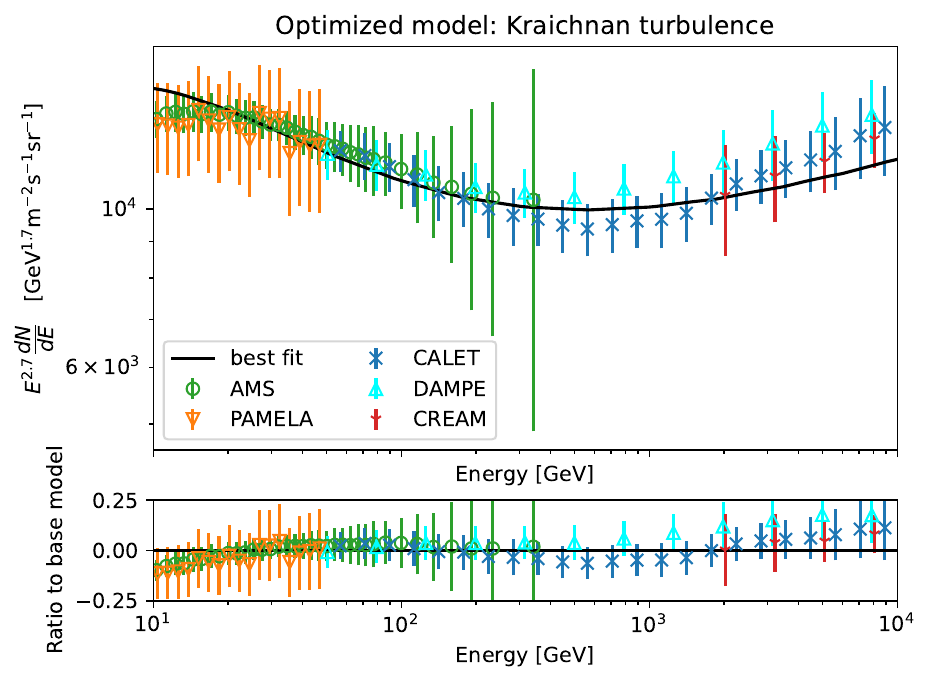}
    \caption{Same as Fig.\ref{fig:f_kolm} but with the best-fit parameters in case of the \textit{Kraichnan} phenomenology.}
    \label{fig:f_kraich}
\end{figure}
Adapting these parameters to our model results in the exemplary cosmic-ray distributions displayed in Fig.\,\ref{fig:f_kolm} and \ref{fig:f_kraich}. 
We find a stronger break in the \textit{Kolmogorov} case, but with an accompanying overshooting of the lower energy part of the cosmic-ray spectrum data since the data sets for the lower energy regime were not considered in our MCMC procedure. Both fits tend to lie beneath the data points at higher energies.
The plotted AMS-02 and PAMELA data have been corrected for the Solar modulation using a force field approximation \citep{Gleeson1968} with the modulation parameter determined by each experiment, respectively. The error bars are given as the combined statistical and systematic errors.
\section{Discussion and conclusions} \label{sec:disccusion}
We carried out an MCMC analysis of a simple model for Galactic cosmic-ray transport, self-consistently including the streaming instability. From this procedure, we obtain optimal parameter ranges that might help to constrain the actual physical values for our Milky Way.
Since we only allowed for the streaming instability to occur, disregarding further possible excitation or damping processes, this model can be considered as an attempt to explore the plausibility of this very effect as a cause of the observed spectral break. Furthermore, since we reduced the complexity to one dimension, the values obtained may be considered as a local value averaged across the Galactic halo. 

The median values in the \textit{Kolmogorov} case suggest a background ion density of $n_i=0.23\,\mathrm{cm}^{-3}$, the case of \textit{Kraichnan} yields $n_i=480\,\mathrm{cm}^{-3}$. In fact, the MCMCs indicate allowed ranges of $n_i\in[0.0025, 7.2]\,\mathrm{cm}^{-3}$ and $n_i\in[23.9,13160]\,\mathrm{cm}^{-3}$, respectively. Values such as the latter are only realized in environments where ionization is far from complete so that ion-neutral damping must then be taken into account and our approach, therefore, is insufficient.
The former value agrees well with, e.g., \citet{Ferrière_1998} who modeled the interstellar hydrogen density and found densities of up to $\sim 0.2\,\mathrm{cm}^{-3}$ for the warm ionized medium. 

In regards to the mean magnetic field $B_0$, both cases do not show clear preferences for a certain value since the marginalized distributions are more or less flat. The suggested values of about $1.8\,\mu$G are, therefore, simply the median of the tested ranges. Local  measurements of the heliosheath flow \citep{Opher2009}, synchrotron emission \citep{Beuermann1985} or $\gamma-$ray data \citep{Strong_2000} indicate values of $4-6\,\mu$G, but also lower strengths around $1\,\mu$G are found from rotation measurements of pulsars \citep{RandLyne1994}.

%
The turbulent component of the magnetic field is afflicted with more uncertainty. Though in spiral galaxies it is assumed to be stronger than the ordered field, some references support lower values for our Galaxy.
\citet{Reichherzer2022} use a turbulence value of $\delta B/B_0\approx 0.5$ in their analysis, which was scaled down due to a likely overestimation, as \citet{Becker-Tjus-Merten-2020} argue.
Here, the turbulence measure $\eta=\delta B^2/B_0^2$ is determined by Eq.\,(\ref{eq:norm_W}) implying that with fixed $B_0$, the variation of $k_0^{-1}$ has a direct effect on the value of $\eta$. The injection scale $k_0^{-1}$ in turn is uncertain and typically estimated to be in a range of $10-100$\,pc \citep[e.g., ][]{Haverkorn_2008, Evoli2018}. 
Our $\eta$-values give us turbulence levels of $\delta B/B_0 =0.091$  and $\delta B/B_0=0.19$ for \textit{Kolmogorov} and \textit{Kraichnan}, respectively. Taking into account the uncertainty of $k_0^{-1}$, our analysis allows ranges of $\delta B/B_0$ of $[0.064,0.2]$ and $[0.13,0.42]$, respectively. Those values are slightly below the values derived by \citet{Reichherzer2022}.

The cosmic-ray injection factor $A$ obtained within this model appears to be of a reasonable order of magnitude, as well. The energy released by supernovae is assumed to be about $10^{51}\,$ergs \citep{Chevalier1977}, of which about 10\% is converted into kinetic energy of protons \citep{Dermer2013}. The Galactic SN rate is estimated to be 1/(36 years) \citep{Kachelriess20251},
1/(40 years) \citep{Tammann1994},
or even between 1/50 and 1/(100 years) \citep{Reed_2005}. Given a Galactic radius $R_d\approx 10\,$kpc, the normalization of $A$ according to Eq.\,(\ref{eq:normalization_A}) would be of the order of $10^{36}\,\mathrm{cm}^{-5}\mathrm{g}^{-3}\mathrm{s}^2$. If we, as a rough estimate, allow for ranges $E_{\mathrm{SN}}\in[5\cdot10^{50},10^{51}]$\,erg, $\xi_{\mathrm{CR}}\in[5,15]\%$, $R_d\in[5,15]\,$kpc and $\mathcal{I}(\alpha)\in [43,248]$ since $\alpha$ ranges from $\sim [4.05,4.3]$, we obtain a range for the injection factor of $A\in[3\cdot 10^{34},2\cdot 10^{37}]\,\mathrm{cm}^{-5}\mathrm{g}^{-3}\mathrm{s}^2$, well consistent with both of our results.

The Kolmogorov constant $c_K$ is associated with some uncertainty and, partly, not used consistently throughout the literature. It commonly appears in the expression relating the wave energy and wavenumber, $W=\tilde{c}_K\epsilon^{2/3}k^{-5/3}$, where $\epsilon = -D_{kk}\nabla W$. \citet{Beresnyak2011} performs MHD simulations and finds $\tilde{c}_K=4.2$, \citet{Yeung1997} yield $\tilde{c}_K$ values of 0.60 and 0.53 for the one- and three-dimensional energy spectra, \citet{Sreenivasan1995} obtains $\tilde{c}_K$=0.53 and a study by \citet{Verma1996} reports $\tilde{c}_K=3.6$. By inserting our diffusion coefficient, Eq. (\ref{eq:D_kk}), and comparing our expression with the ansatz in which $W\propto \tilde{c}_K$, we assume $c_K=(\tilde{c}_K)^{-3/2}$, so that our result for the \textit{Kolmogorov} case, $c_K\approx 0.37$ is in the expected order of magnitude. In case of a \textit{Kraichnan}-like cascade, numerical simulations were carried out by \citet{Biskamp1989} which indicate $\tilde{c}_K=1.8\pm0.2$. \citet{MatthaeusZhou1989} suggest a relation of the Kolmogorov constant and the 'Kraichnan constant' and  conclude $\tilde{c}_K=1.22-1.87$. Since in this case, the relation is $W\propto \tilde{c}_Kk^{-3/2}$, our constant would be calculated as $c_K=(\tilde{c}_K)^{-2}.$  The best-fit result from the MCMC analysis yields $c_K\approx 1.41$, corresponding to $\tilde{c}_K=1.99$.

In both turbulence cases, the break at a few hundred GeV is resembled, the stronger softening of the \textit{Kolmogorov} spectrum is in better accordance with other data such as DAMPE. The overshooting in the lower energy regime might allude to a spectral break in the source distribution. 

So, in summary, our MCMC analysis shows that the streaming-instability in combination with a \textit{Kolmogorov}- as well as \textit{Kraichnan}-like cascade is capable of self-consistently causing a spectral break that resembles the one emerging at a few hundred GeV in available cosmic ray data. The associated values for the various model parameters favor the former phenomenology,

\section*{Acknowledgements}
This work was supported by the Deutsche Forschungsgemeinschaft within the Collaborative Research Center SFB1491 (project no.\ 445052434).
\bibliography{literature.bib}{}

@ARTICLE{biermann2010,
       author = {{Biermann}, Peter L. and {Becker}, Julia K. and {Dreyer}, Jens and {Meli}, Athina and {Seo}, Eun-Suk and {Stanev}, Todor},
title = "{The Origin of Cosmic Rays: Explosions of Massive Stars with Magnetic Winds and Their Supernova Mechanism}",
      journal = {\apj},
         year = 2010,
        month = dec,
       volume = {725},
       number = {1},
        pages = {184-187},
          doi = {10.1088/0004-637X/725/1/184},
archivePrefix = {arXiv},
       eprint = {1009.5592},
 primaryClass = {astro-ph.HE},
       adsurl = {https://ui.adsabs.harvard.edu/abs/2010ApJ...725..184B}
}

@article{Aloisio,
doi = {10.1088/1475-7516/2013/07/001},
url = {https://dx.doi.org/10.1088/1475-7516/2013/07/001},
year = {2013},
month = {jul},
publisher = {},
volume = {2013},
number = {07},
pages = {001},
author = {Roberto Aloisio and  Pasquale Blasi},
title = {Propagation of galactic cosmic rays in the presence of self-generated turbulence},
journal = {Journal of Cosmology and Astroparticle Physics}
}

@ARTICLE{blasi2012,
       author = {{Blasi}, Pasquale and {Amato}, Elena and {Serpico}, Pasquale D.},
        title = "{Spectral Breaks as a Signature of Cosmic Ray Induced Turbulence in the Galaxy}",
      journal = {
    Physical Review Letters},
         year = 2012,
        month = aug,
       volume = {109},
       number = {6},
          eid = {061101},
        pages = {061101},
archivePrefix = {arXiv},
       eprint = {1207.3706},
 primaryClass = {astro-ph.HE}
}

@article{leith,
    author = {Leith, C. E.},
    title = {Diffusion Approximation to Inertial Energy Transfer in Isotropic Turbulence},
    journal = {The Physics of Fluids},
    volume = {10},
    number = {7},
    pages = {1409-1416},
    year = {1967},
    month = {07},
    issn = {0031-9171},
    doi = {10.1063/1.1762300},
    url = {https://doi.org/10.1063/1.1762300},
    eprint = {https://pubs.aip.org/aip/pfl/article-pdf/10/7/1409/12451068/1409\_1\_online.pdf},
}

@article{zhou1990,
author = {Zhou, Ye and Matthaeus, William H.},
title = {Models of inertial range spectra of interplanetary magnetohydrodynamic turbulence},
journal = {Journal of Geophysical Research: Space Physics},
volume = {95},
number = {A9},
pages = {14881-14892},
doi = {https://doi.org/10.1029/JA095iA09p14881},
url = {https://agupubs.onlinelibrary.wiley.com/doi/abs/10.1029/JA095iA09p14881},
eprint = {https://agupubs.onlinelibrary.wiley.com/doi/pdf/10.1029/JA095iA09p14881},
year = {1990}
}

@ARTICLE{miller1995,
       author = {{Miller}, James A. and {Roberts}, D. Aaron},
        title = "{Stochastic Proton Acceleration by Cascading Alfven Waves in Impulsive Solar Flares}",
      journal = {\apj},
         year = 1995,
        month = oct,
       volume = {452},
        pages = {912},
          doi = {10.1086/176359},
       adsurl = {https://ui.adsabs.harvard.edu/abs/1995ApJ...452..912M}
}

@ARTICLE{skilling1971,
       author = {{Skilling}, John},
        title = "{Cosmic Rays in the Galaxy: Convection or Diffusion?}",
      journal = {\apj},
         year = 1971,
        month = dec,
       volume = {170},
        pages = {265},
          doi = {10.1086/151210},
       adsurl = {https://ui.adsabs.harvard.edu/abs/1971ApJ...170..265S}
}

@ARTICLE{mcmc,
       author = {{Foreman-Mackey}, Daniel and {Hogg}, David W. and {Lang}, Dustin and {Goodman}, Jonathan},
        title = "{emcee: The MCMC Hammer}",
      journal = {\pasp},
         year = 2013,
        month = mar,
       volume = {125},
       number = {925},
        pages = {306},
          doi = {10.1086/670067},
archivePrefix = {arXiv},
       eprint = {1202.3665},
 primaryClass = {astro-ph.IM},
       adsurl = {https://ui.adsabs.harvard.edu/abs/2013PASP..125..306F}
}

@ARTICLE{likelihood,
       author = {{Hogg}, David W. and {Bovy}, Jo and {Lang}, Dustin},
        title = "{Data analysis recipes: Fitting a model to data}",
      journal = {arXiv e-prints},
         year = 2010,
        month = aug,
          eid = {arXiv:1008.4686},
        pages = {arXiv:1008.4686},
          doi = {10.48550/arXiv.1008.4686},
archivePrefix = {arXiv},
       eprint = {1008.4686},
 primaryClass = {astro-ph.IM},
       adsurl = {https://ui.adsabs.harvard.edu/abs/2010arXiv1008.4686H}
}

@ARTICLE{Kulsrud1969,
       author = {{Kulsrud}, Russell and {Pearce}, William P.},
        title = "{The Effect of Wave-Particle Interactions on the Propagation of Cosmic Rays}",
      journal = {\apj},
         year = 1969,
        month = may,
       volume = {156},
        pages = {445},
          doi = {10.1086/149981},
       adsurl = {https://ui.adsabs.harvard.edu/abs/1969ApJ...156..445K}
}

@ARTICLE{scipy,
  author  = {Virtanen, Pauli and Gommers, Ralf and Oliphant, Travis E. and
            Haberland, Matt and Reddy, Tyler and Cournapeau, David and
            Burovski, Evgeni and Peterson, Pearu and Weckesser, Warren and
            Bright, Jonathan and {van der Walt}, St{\'e}fan J. and
            Brett, Matthew and Wilson, Joshua and Millman, K. Jarrod and
            Mayorov, Nikolay and Nelson, Andrew R. J. and Jones, Eric and
            Kern, Robert and Larson, Eric and Carey, C J and
            Polat, {\.I}lhan and Feng, Yu and Moore, Eric W. and
            {VanderPlas}, Jake and Laxalde, Denis and Perktold, Josef and
            Cimrman, Robert and Henriksen, Ian and Quintero, E. A. and
            Harris, Charles R. and Archibald, Anne M. and
            Ribeiro, Ant{\^o}nio H. and Pedregosa, Fabian and
            {van Mulbregt}, Paul and {SciPy 1.0 Contributors}},
  title   = {{{SciPy} 1.0: Fundamental Algorithms for Scientific
            Computing in Python}},
  journal = {Nature Methods},
  year    = {2020},
  volume  = {17},
  pages   = {261--272},
  adsurl  = {https://rdcu.be/b08Wh},
  doi     = {10.1038/s41592-019-0686-2},
}

@ARTICLE{CREAM,
       author = {{Yoon}, Y.~S. and {Ahn}, H.~S. and {Allison}, P.~S. and {Bagliesi}, M.~G. and {Beatty}, J.~J. and {Bigongiari}, G. and {Boyle}, P.~J. and {Childers}, J.~T. and {Conklin}, N.~B. and {Coutu}, S. and {DuVernois}, M.~A. and {Ganel}, O. and {Han}, J.~H. and {Jeon}, J.~A. and {Kim}, K.~C. and {Lee}, M.~H. and {Lutz}, L. and {Maestro}, P. and {Malinine}, A. and {Marrocchesi}, P.~S. and {Minnick}, S.~A. and {Mognet}, S.~I. and {Nam}, S. and {Nutter}, S. and {Park}, I.~H. and {Park}, N.~H. and {Seo}, E.~S. and {Sina}, R. and {Swordy}, S. and {Wakely}, S.~P. and {Wu}, J. and {Yang}, J. and {Zei}, R. and {Zinn}, S.~Y.},
        title = "{Cosmic-ray Proton and Helium Spectra from the First CREAM Flight}",
      journal = {\apj},
         year = 2011,
        month = feb,
       volume = {728},
       number = {2},
          eid = {122},
        pages = {122},
          doi = {10.1088/0004-637X/728/2/122},
archivePrefix = {arXiv},
       eprint = {1102.2575},
 primaryClass = {astro-ph.HE},
       adsurl = {https://ui.adsabs.harvard.edu/abs/2011ApJ...728..122Y}
}

@article{BELL2013,
title = {Cosmic ray acceleration},
journal = {Astroparticle Physics},
volume = {43},
pages = {56-70},
year = {2013},
note = {Seeing the High-Energy Universe with the Cherenkov Telescope Array - The Science Explored with the CTA},
issn = {0927-6505},
doi = {https://doi.org/10.1016/j.astropartphys.2012.05.022},
url = {https://www.sciencedirect.com/science/article/pii/S0927650512001272},
author = {A.R. Bell}
}

@ARTICLE{Reichherzer2022,
       author = {{Reichherzer}, P. and {Merten}, L. and {D{\"o}rner}, J. and {Becker Tjus}, J. and {Pueschel}, M.~J. and {Zweibel}, E.~G.},
        title = "{Regimes of cosmic-ray diffusion in Galactic turbulence}",
      journal = {SN Applied Sciences},
         year = 2022,
        month = jan,
       volume = {4},
          eid = {15},
        pages = {15},
          doi = {10.1007/s42452-021-04891-z},
archivePrefix = {arXiv},
       eprint = {2104.13093},
 primaryClass = {astro-ph.HE},
       adsurl = {https://ui.adsabs.harvard.edu/abs/2022SNAS....4...15R}
}

@article{SHALCHI2009,
title = {Diffusive shock acceleration in supernova remnants: On the validity of the Bohm limit},
journal = {Astroparticle Physics},
volume = {31},
number = {3},
pages = {237-242},
year = {2009},
issn = {0927-6505},
doi = {https://doi.org/10.1016/j.astropartphys.2009.01.007},
url = {https://www.sciencedirect.com/science/article/pii/S0927650509000280},
author = {A. Shalchi}
}

@ARTICLE{DAMPE,
       author = {{An}, Q. and {Asfandiyarov}, R. and {Azzarello}, P. and {Bernardini}, P. and {Bi}, X.~J. and {Cai}, M.~S. and {Chang}, J. and {Chen}, D.~Y. and {Chen}, H.~F. and {Chen}, J.~L. and {Chen}, W. and {Cui}, M.~Y. and {Cui}, T.~S. and {Dai}, H.~T. and {D'Amone}, A. and {De Benedittis}, A. and {De Mitri}, I. and {Di Santo}, M. and {Ding}, M. and {Dong}, T.~K. and {Dong}, Y.~F. and {Dong}, Z.~X. and {Donvito}, G. and {Droz}, D. and {Duan}, J.~L. and {Duan}, K.~K. and {D'Urso}, D. and {Fan}, R.~R. and {Fan}, Y.~Z. and {Fang}, F. and {Feng}, C.~Q. and {Feng}, L. and {Fusco}, P. and {Gallo}, V. and {Gan}, F.~J. and {Gao}, M. and {Gargano}, F. and {Gong}, K. and {Gong}, Y.~Z. and {Guo}, D.~Y. and {Guo}, J.~H. and {Guo}, X.~L. and {Han}, S.~X. and {Hu}, Y.~M. and {Huang}, G.~S. and {Huang}, X.~Y. and {Huang}, Y.~Y. and {Ionica}, M. and {Jiang}, W. and {Jin}, X. and {Kong}, J. and {Lei}, S.~J. and {Li}, S. and {Li}, W.~L. and {Li}, X. and {Li}, X.~Q. and {Li}, Y. and {Liang}, Y.~F. and {Liang}, Y.~M. and {Liao}, N.~H. and {Liu}, C.~M. and {Liu}, H. and {Liu}, J. and {Liu}, S.~B. and {Liu}, W.~Q. and {Liu}, Y. and {Loparco}, F. and {Luo}, C.~N. and {Ma}, M. and {Ma}, P.~X. and {Ma}, S.~Y. and {Ma}, T. and {Ma}, X.~Y. and {Marsella}, G. and {Mazziotta}, M.~N. and {Mo}, D. and {Niu}, X.~Y. and {Pan}, X. and {Peng}, W.~X. and {Peng}, X.~Y. and {Qiao}, R. and {Rao}, J.~N. and {Salinas}, M.~M. and {Shang}, G.~Z. and {Shen}, W.~H. and {Shen}, Z.~Q. and {Shen}, Z.~T. and {Song}, J.~X. and {Su}, H. and {Su}, M. and {Sun}, Z.~Y. and {Surdo}, A. and {Teng}, X.~J. and {Tykhonov}, A. and {Vitillo}, S. and {Wang}, C. and {Wang}, H. and {Wang}, H.~Y. and {Wang}, J.~Z. and {Wang}, L.~G. and {Wang}, Q. and {Wang}, S. and {Wang}, X.~H. and {Wang}, X.~L. and {Wang}, Y.~F. and {Wang}, Y.~P. and {Wang}, Y.~Z. and {Wang}, Z.~M. and {Wei}, D.~M. and {Wei}, J.~J. and {Wei}, Y.~F. and {Wen}, S.~C. and {Wu}, D. and {Wu}, J. and {Wu}, L.~B. and {Wu}, S.~S. and {Wu}, X. and {Xi}, K. and {Xia}, Z.~Q. and {Xu}, H.~T. and {Xu}, Z.~H. and {Xu}, Z.~L. and {Xu}, Z.~Z. and {Xue}, G.~F. and {Yang}, H.~B. and {Yang}, P. and {Yang}, Y.~Q. and {Yang}, Z.~L. and {Yao}, H.~J. and {Yu}, Y.~H. and {Yuan}, Q. and {Yue}, C. and {Zang}, J.~J. and {Zhang}, F. and {Zhang}, J.~Y. and {Zhang}, J.~Z. and {Zhang}, P.~F. and {Zhang}, S.~X. and {Zhang}, W.~Z. and {Zhang}, Y. and {Zhang}, Y.~J. and {Zhang}, Y.~L. and {Zhang}, Y.~P. and {Zhang}, Y.~Q. and {Zhang}, Z. and {Zhang}, Z.~Y. and {Zhao}, H. and {Zhao}, H.~Y. and {Zhao}, X.~F. and {Zhou}, C.~Y. and {Zhou}, Y. and {Zhu}, X. and {Zhu}, Y. and {Zimmer}, S.},
        title = "{Measurement of the cosmic ray proton spectrum from 40 GeV to 100 TeV with the DAMPE satellite}",
      journal = {Science Advances},
         year = 2019,
        month = sep,
       volume = {5},
       number = {9},
          eid = {eaax3793},
        pages = {eaax3793},
          doi = {10.1126/sciadv.aax3793},
archivePrefix = {arXiv},
       eprint = {1909.12860},
 primaryClass = {astro-ph.HE},
       adsurl = {https://ui.adsabs.harvard.edu/abs/2019SciA....5.3793A}
}

@ARTICLE{CALET,
       author = {{Adriani}, O. and {Akaike}, Y. and {Asano}, K. and {Asaoka}, Y. and {Berti}, E. and {Bigongiari}, G. and {Binns}, W.~R. and {Bongi}, M. and {Brogi}, P. and {Bruno}, A. and {Buckley}, J.~H. and {Cannady}, N. and {Castellini}, G. and {Checchia}, C. and {Cherry}, M.~L. and {Collazuol}, G. and {Ebisawa}, K. and {Ficklin}, A.~W. and {Fuke}, H. and {Gonzi}, S. and {Guzik}, T.~G. and {Hams}, T. and {Hibino}, K. and {Ichimura}, M. and {Ioka}, K. and {Ishizaki}, W. and {Israel}, M.~H. and {Kasahara}, K. and {Kataoka}, J. and {Kataoka}, R. and {Katayose}, Y. and {Kato}, C. and {Kawanaka}, N. and {Kawakubo}, Y. and {Kobayashi}, K. and {Kohri}, K. and {Krawczynski}, H.~S. and {Krizmanic}, J.~F. and {Maestro}, P. and {Marrocchesi}, P.~S. and {Messineo}, A.~M. and {Mitchell}, J.~W. and {Miyake}, S. and {Moiseev}, A.~A. and {Mori}, M. and {Mori}, N. and {Motz}, H.~M. and {Munakata}, K. and {Nakahira}, S. and {Nishimura}, J. and {de Nolfo}, G.~A. and {Okuno}, S. and {Ormes}, J.~F. and {Ozawa}, S. and {Pacini}, L. and {Papini}, P. and {Rauch}, B.~F. and {Ricciarini}, S.~B. and {Sakai}, K. and {Sakamoto}, T. and {Sasaki}, M. and {Shimizu}, Y. and {Shiomi}, A. and {Spillantini}, P. and {Stolzi}, F. and {Sugita}, S. and {Sulaj}, A. and {Takita}, M. and {Tamura}, T. and {Terasawa}, T. and {Torii}, S. and {Tsunesada}, Y. and {Uchihori}, Y. and {Vannuccini}, E. and {Wefel}, J.~P. and {Yamaoka}, K. and {Yanagita}, S. and {Yoshida}, A. and {Yoshida}, K. and {Zober}, W.~V. and {Calet Collaboration}},
        title = "{Observation of Spectral Structures in the Flux of Cosmic-Ray Protons from 50 GeV to 60 TeV with the Calorimetric Electron Telescope on the International Space Station}",
      journal = {\prl},
         year = 2022,
        month = sep,
       volume = {129},
       number = {10},
          eid = {101102},
        pages = {101102},
          doi = {10.1103/PhysRevLett.129.101102},
archivePrefix = {arXiv},
       eprint = {2209.01302},
 primaryClass = {astro-ph.HE},
       adsurl = {https://ui.adsabs.harvard.edu/abs/2022PhRvL.129j1102A}
}

@ARTICLE{PAMELA,
       author = {{Adriani}, O. and {Barbarino}, G.~C. and {Bazilevskaya}, G.~A. and {Bellotti}, R. and {Boezio}, M. and {Bogomolov}, E.~A. and {Bonechi}, L. and {Bongi}, M. and {Bonvicini}, V. and {Borisov}, S. and {Bottai}, S. and {Bruno}, A. and {Cafagna}, F. and {Campana}, D. and {Carbone}, R. and {Carlson}, P. and {Casolino}, M. and {Castellini}, G. and {Consiglio}, L. and {De Pascale}, M.~P. and {De Santis}, C. and {De Simone}, N. and {Di Felice}, V. and {Galper}, A.~M. and {Gillard}, W. and {Grishantseva}, L. and {Jerse}, G. and {Karelin}, A.~V. and {Koldashov}, S.~V. and {Krutkov}, S.~Y. and {Kvashnin}, A.~N. and {Leonov}, A. and {Malakhov}, V. and {Malvezzi}, V. and {Marcelli}, L. and {Mayorov}, A.~G. and {Menn}, W. and {Mikhailov}, V.~V. and {Mocchiutti}, E. and {Monaco}, A. and {Mori}, N. and {Nikonov}, N. and {Osteria}, G. and {Palma}, F. and {Papini}, P. and {Pearce}, M. and {Picozza}, P. and {Pizzolotto}, C. and {Ricci}, M. and {Ricciarini}, S.~B. and {Rossetto}, L. and {Sarkar}, R. and {Simon}, M. and {Sparvoli}, R. and {Spillantini}, P. and {Stozhkov}, Y.~I. and {Vacchi}, A. and {Vannuccini}, E. and {Vasilyev}, G. and {Voronov}, S.~A. and {Yurkin}, Y.~T. and {Wu}, J. and {Zampa}, G. and {Zampa}, N. and {Zverev}, V.~G.},
        title = "{PAMELA Measurements of Cosmic-Ray Proton and Helium Spectra}",
      journal = {Science},
         year = 2011,
        month = apr,
       volume = {332},
       number = {6025},
        pages = {69},
          doi = {10.1126/science.1199172},
archivePrefix = {arXiv},
       eprint = {1103.4055},
 primaryClass = {astro-ph.HE},
       adsurl = {https://ui.adsabs.harvard.edu/abs/2011Sci...332...69A}
}

@ARTICLE{AMS,
       author = {{Aguilar}, M. and {Ali Cavasonza}, L. and {Ambrosi}, G. and {Arruda}, L. and {Attig}, N. and {Barao}, F. and {Barrin}, L. and {Bartoloni}, A. and {Ba{\c{s}}e{\u{g}}mez-du Pree}, S. and {Bates}, J. and {Battiston}, R. and {Behlmann}, M. and {Beischer}, B. and {Berdugo}, J. and {Bertucci}, B. and {Bindi}, V. and {de Boer}, W. and {Bollweg}, K. and {Borgia}, B. and {Boschini}, M.~J. and {Bourquin}, M. and {Bueno}, E.~F. and {Burger}, J. and {Burger}, W.~J. and {Burmeister}, S. and {Cai}, X.~D. and {Capell}, M. and {Casaus}, J. and {Castellini}, G. and {Cervelli}, F. and {Chang}, Y.~H. and {Chen}, G.~M. and {Chen}, H.~S. and {Chen}, Y. and {Cheng}, L. and {Chou}, H.~Y. and {Chouridou}, S. and {Choutko}, V. and {Chung}, C.~H. and {Clark}, C. and {Coignet}, G. and {Consolandi}, C. and {Contin}, A. and {Corti}, C. and {Cui}, Z. and {Dadzie}, K. and {Dai}, Y.~M. and {Delgado}, C. and {Della Torre}, S. and {Demirk{\"o}z}, M.~B. and {Derome}, L. and {Di Falco}, S. and {Di Felice}, V. and {D{\'\i}az}, C. and {Dimiccoli}, F. and {von Doetinchem}, P. and {Dong}, F. and {Donnini}, F. and {Duranti}, M. and {Egorov}, A. and {Eline}, A. and {Feng}, J. and {Fiandrini}, E. and {Fisher}, P. and {Formato}, V. and {Freeman}, C. and {Galaktionov}, Y. and {G{\'a}mez}, C. and {Garc{\'\i}a-L{\'o}pez}, R.~J. and {Gargiulo}, C. and {Gast}, H. and {Gebauer}, I. and {Gervasi}, M. and {Giovacchini}, F. and {G{\'o}mez-Coral}, D.~M. and {Gong}, J. and {Goy}, C. and {Grabski}, V. and {Grandi}, D. and {Graziani}, M. and {Guo}, K.~H. and {Haino}, S. and {Han}, K.~C. and {Hashmani}, R.~K. and {He}, Z.~H. and {Heber}, B. and {Hsieh}, T.~H. and {Hu}, J.~Y. and {Huang}, Z.~C. and {Hungerford}, W. and {Incagli}, M. and {Jang}, W.~Y. and {Jia}, Yi and {Jinchi}, H. and {Kanishev}, K. and {Khiali}, B. and {Kim}, G.~N. and {Kirn}, Th. and {Konyushikhin}, M. and {Kounina}, O. and {Kounine}, A. and {Koutsenko}, V. and {Kuhlman}, A. and {Kulemzin}, A. and {La Vacca}, G. and {Laudi}, E. and {Laurenti}, G. and {Lazzizzera}, I. and {Lebedev}, A. and {Lee}, H.~T. and {Lee}, S.~C. and {Leluc}, C. and {Li}, J.~Q. and {Li}, M. and {Li}, Q. and {Li}, S. and {Li}, T.~X. and {Li}, Z.~H. and {Light}, C. and {Lin}, C.~H. and {Lippert}, T. and {Liu}, Z. and {Lu}, S.~Q. and {Lu}, Y.~S. and {Luebelsmeyer}, K. and {Luo}, J.~Z. and {Lyu}, S.~S. and {Machate}, F. and {Ma{\~n}{\'a}}, C. and {Mar{\'\i}n}, J. and {Marquardt}, J. and {Martin}, T. and {Mart{\'\i}nez}, G. and {Masi}, N. and {Maurin}, D. and {Menchaca-Rocha}, A. and {Meng}, Q. and {Mo}, D.~C. and {Molero}, M. and {Mott}, P. and {Mussolin}, L. and {Ni}, J.~Q. and {Nikonov}, N. and {Nozzoli}, F. and {Oliva}, A. and {Orcinha}, M. and {Palermo}, M. and {Palmonari}, F. and {Paniccia}, M. and {Pashnin}, A. and {Pauluzzi}, M. and {Pensotti}, S. and {Phan}, H.~D. and {Plyaskin}, V. and {Pohl}, M. and {Porter}, S. and {Qi}, X.~M. and {Qin}, X. and {Qu}, Z.~Y. and {Quadrani}, L. and {Rancoita}, P.~G. and {Rapin}, D. and {Reina Conde}, A. and {Rosier-Lees}, S. and {Rozhkov}, A. and {Rozza}, D. and {Sagdeev}, R. and {Schael}, S. and {Schmidt}, S.~M. and {Schulz von Dratzig}, A. and {Schwering}, G. and {Seo}, E.~S. and {Shan}, B.~S. and {Shi}, J.~Y. and {Siedenburg}, T. and {Solano}, C. and {Song}, J.~W. and {Sonnabend}, R. and {Sun}, Q. and {Sun}, Z.~T. and {Tacconi}, M. and {Tang}, X.~W. and {Tang}, Z.~C. and {Tian}, J. and {Ting}, Samuel C.~C. and {Ting}, S.~M. and {Tomassetti}, N. and {Torsti}, J. and {T{\"u}ys{\"u}z}, C. and {Urban}, T. and {Usoskin}, I. and {Vagelli}, V. and {Vainio}, R. and {Valente}, E. and {Valtonen}, E. and {V{\'a}zquez Acosta}, M. and {Vecchi}, M. and {Velasco}, M. and {Vialle}, J.~P. and {Wang}, L.~Q.},
        title = "{The Alpha Magnetic Spectrometer (AMS) on the international space station: Part II - Results from the first seven years}",
      journal = {\physrep},
         year = 2021,
        month = feb,
       volume = {894},
        pages = {1-116},
          doi = {10.1016/j.physrep.2020.09.003},
       adsurl = {https://ui.adsabs.harvard.edu/abs/2021PhR...894....1A}
}

@article{cornerplot, doi = {10.21105/joss.00024}, url = {https://doi.org/10.21105/joss.00024}, year = {2016}, publisher = {The Open Journal}, volume = {1}, number = {2}, pages = {24}, author = {Daniel Foreman-Mackey}, title = {corner.py: Scatterplot matrices in Python}, journal = {Journal of Open Source Software} }

@article{Silver_2024,
doi = {10.3847/1538-4357/ad1ce8},
url = {https://dx.doi.org/10.3847/1538-4357/ad1ce8},
year = {2024},
month = {mar},
publisher = {The American Astronomical Society},
volume = {963},
number = {2},
pages = {111},
author = {Silver, Ethan and Orlando, Elena},
title = {Testing Cosmic-Ray Propagation Scenarios with AMS-02 and Voyager Data},
journal = {The Astrophysical Journal}
}

@article{Pan_2023,
doi = {10.1088/1674-4527/acf443},
url = {https://dx.doi.org/10.1088/1674-4527/acf443},
year = {2023},
month = {oct},
publisher = {National Astromonical Observatories, CAS and IOP Publishing},
volume = {23},
number = {11},
pages = {115002},
author = {Pan, Xu and Yuan, Qiang},
title = {Injection Spectra of Different Species of Cosmic Rays from AMS-02, ACE-CRIS and Voyager-1},
journal = {Research in Astronomy and Astrophysics}
}

@ARTICLE{Tomassetti2012,
       author = {{Tomassetti}, Nicola},
        title = "{Origin of the Cosmic-Ray Spectral Hardening}",
      journal = {\apjl},
         year = 2012,
        month = jun,
       volume = {752},
       number = {1},
          eid = {L13},
        pages = {L13},
          doi = {10.1088/2041-8205/752/1/L13},
archivePrefix = {arXiv},
       eprint = {1204.4492},
 primaryClass = {astro-ph.HE},
       adsurl = {https://ui.adsabs.harvard.edu/abs/2012ApJ...752L..13T}
}

@ARTICLE{Liu2018,
       author = {{Liu}, Wei and {Yao}, Yu-hua and {Guo}, Yi-Qing},
        title = "{Revisiting the Spatially Dependent Propagation Model with the Latest Observations of Cosmic-Ray Nuclei}",
      journal = {\apj},
         year = 2018,
        month = dec,
       volume = {869},
       number = {2},
          eid = {176},
        pages = {176},
          doi = {10.3847/1538-4357/aaef39},
archivePrefix = {arXiv},
       eprint = {1802.03602},
 primaryClass = {astro-ph.HE},
       adsurl = {https://ui.adsabs.harvard.edu/abs/2018ApJ...869..176L}
}

@article{Ferrière_1998,
doi = {10.1086/305469},
url = {https://dx.doi.org/10.1086/305469},
year = {1998},
month = {apr},
publisher = {},
volume = {497},
number = {2},
pages = {759},
author = {Ferrière, Katia},
title = {Global Model of the Interstellar Medium in our Galaxy with New Constraints on the Hot Gas Component},
journal = {The Astrophysical Journal}
}

@article{Kachelriess20251,
title = {Galactic distribution of supernovae and OB associations},
journal = {Computer Physics Communications},
volume = {311},
pages = {109537},
year = {2025},
issn = {0010-4655},
doi = {https://doi.org/10.1016/j.cpc.2025.109537},
url = {https://www.sciencedirect.com/science/article/pii/S0010465525000402},
author = {M. Kachelrieß and V. Mikalsen}
}

@ARTICLE{Beuermann1985,
       author = {{Beuermann}, K. and {Kanbach}, G. and {Berkhuijsen}, E.~M.},
        title = "{Radio structure of the Galaxy: thick disk and thin disk at 408 MHz.}",
      journal = {\aap},
         year = 1985,
        month = dec,
       volume = {153},
        pages = {17-34},
       adsurl = {https://ui.adsabs.harvard.edu/abs/1985A&A...153...17B}
}

@ARTICLE{Chevalier1977,
       author = {{Chevalier}, R.~A.},
        title = "{The interaction of supernovae with the interstellar medium.}",
      journal = {\araa},
         year = 1977,
        month = jan,
       volume = {15},
        pages = {175-196},
          doi = {10.1146/annurev.aa.15.090177.001135},
       adsurl = {https://ui.adsabs.harvard.edu/abs/1977ARA&A..15..175C}
}

@article{Strong_2000,
doi = {10.1086/309038},
url = {https://dx.doi.org/10.1086/309038},
year = {2000},
month = {jul},
publisher = {},
volume = {537},
number = {2},
pages = {763},
author = {Strong, Andrew W. and Moskalenko, Igor V. and Reimer, Olaf},
title = {Diffuse Continuum Gamma Rays from the Galaxy},
journal = {The Astrophysical Journal}
}

@article{RandLyne1994,
    author = {Rand, Richard. J. and Lyne, A. G.},
    title = {New rotation measures of distant pulsars in the inner Galaxy and magnetic field reversals},
    journal = {Monthly Notices of the Royal Astronomical Society},
    volume = {268},
    number = {2},
    pages = {497-505},
    year = {1994},
    month = {05},
    issn = {0035-8711},
    doi = {10.1093/mnras/268.2.497},
    url = {https://doi.org/10.1093/mnras/268.2.497},
    eprint = {https://academic.oup.com/mnras/article-pdf/268/2/497/3591223/mnras268-0497.pdf},
}

@article{Reed_2005,
doi = {10.1086/444474},
url = {https://dx.doi.org/10.1086/444474},
year = {2005},
month = {oct},
publisher = {},
volume = {130},
number = {4},
pages = {1652},
author = {Reed, B. Cameron},
title = {New Estimates of the Solar-Neighborhood Massive Star Birthrate and the Galactic Supernova Rate},
journal = {The Astronomical Journal}
}

@ARTICLE{Tammann1994,
       author = {{Tammann}, G.~A. and {Loeffler}, W. and {Schroeder}, A.},
        title = "{The Galactic Supernova Rate}",
      journal = {\apjs},
         year = 1994,
        month = jun,
       volume = {92},
        pages = {487},
          doi = {10.1086/192002},
       adsurl = {https://ui.adsabs.harvard.edu/abs/1994ApJS...92..487T}
}

@ARTICLE{Dermer2013,
       author = {{Dermer}, C.~D. and {Powale}, G.},
        title = "{Gamma rays from cosmic rays in supernova remnants}",
      journal = {\aap},
         year = 2013,
        month = may,
       volume = {553},
          eid = {A34},
        pages = {A34},
          doi = {10.1051/0004-6361/201220394},
archivePrefix = {arXiv},
       eprint = {1210.8071},
 primaryClass = {astro-ph.HE},
       adsurl = {https://ui.adsabs.harvard.edu/abs/2013A&A...553A..34D}
}

@ARTICLE{Opher2009,
       author = {{Opher}, M. and {Bibi}, F. Alouani and {Toth}, G. and {Richardson}, J.~D. and {Izmodenov}, V.~V. and {Gombosi}, T.~I.},
        title = "{A strong, highly-tilted interstellar magnetic field near the Solar System}",
      journal = {\nat},
         year = 2009,
        month = dec,
       volume = {462},
       number = {7276},
        pages = {1036-1038},
          doi = {10.1038/nature08567},
       adsurl = {https://ui.adsabs.harvard.edu/abs/2009Natur.462.1036O}
}

@ARTICLE{Greisen-1966,
       author = {{Greisen}, Kenneth},
        title = "{End to the Cosmic-Ray Spectrum?}",
      journal = {\prl},
         year = 1966,
        month = apr,
       volume = {16},
       number = {17},
        pages = {748-750},
          doi = {10.1103/PhysRevLett.16.748},
       adsurl = {https://ui.adsabs.harvard.edu/abs/1966PhRvL..16..748G}
}

@ARTICLE{Zatsepin-Kuzmin-1966,
       author = {{Zatsepin}, G.~T. and {Kuz'min}, V.~A.},
        title = "{Upper Limit of the Spectrum of Cosmic Rays}",
      journal = {Soviet Journal of Experimental and Theoretical Physics Letters},
         year = 1966,
        month = aug,
       volume = {4},
        pages = {78},
       adsurl = {https://ui.adsabs.harvard.edu/abs/1966JETPL...4...78Z}
}

@ARTICLE{Potgieter-1998,
       author = {{Potgieter}, M.~S.},
        title = "{The Modulation of Galactic Cosmic Rays in the Heliosphere: Theory and Models}",
      journal = {\ssr},
         year = 1998,
        month = jan,
       volume = {83},
        pages = {147-158},
       adsurl = {https://ui.adsabs.harvard.edu/abs/1998SSRv...83..147P}
}

@ARTICLE{Becker-Tjus-Merten-2020,
       author = {{Becker Tjus}, Julia and {Merten}, Lukas},
        title = "{Closing in on the origin of Galactic cosmic rays using multimessenger information}",
      journal = {\physrep},
         year = 2020,
        month = aug,
       volume = {872},
        pages = {1-98},
          doi = {10.1016/j.physrep.2020.05.002},
archivePrefix = {arXiv},
       eprint = {2002.00964},
 primaryClass = {astro-ph.HE},
       adsurl = {https://ui.adsabs.harvard.edu/abs/2020PhR...872....1B}
}

@ARTICLE{Kampert-2013,
       author = {{Kampert}, Karl-Heinz},
        title = "{Ultrahigh-Energy Cosmic Rays: Results and Prospects}",
      journal = {Brazilian Journal of Physics},
         year = 2013,
        month = dec,
       volume = {43},
       number = {5-6},
        pages = {375-382},
          doi = {10.1007/s13538-013-0150-1},
archivePrefix = {arXiv},
       eprint = {1305.2363},
 primaryClass = {astro-ph.HE},
       adsurl = {https://ui.adsabs.harvard.edu/abs/2013BrJPh..43..375K}
}

@ARTICLE{Abreu-etal-2021,
       author = {{Abreu}, P. and {Aglietta}, M. and {Albury}, J.~M. and {Allekotte}, I. and {Almela}, A. and {Alvarez-Mu{\~n}iz}, J. and {Alves Batista}, R. and {Anastasi}, G.~A. and {Anchordoqui}, L. and {Andrada}, B. and {Andringa}, S. and {Aramo}, C. and {Ara{\'u}jo Ferreira}, P.~R. and {Arteaga Vel{\'a}zquez}, J.~C. and {Asorey}, H. and {Assis}, P. and {Avila}, G. and {Badescu}, A.~M. and {Bakalova}, A. and {Balaceanu}, A. and {Barbato}, F. and {Barreira Luz}, R.~J. and {Becker}, K.~H. and {Bellido}, J.~A. and {Berat}, C. and {Bertaina}, M.~E. and {Bertou}, X. and {Biermann}, P.~L. and {Billoir}, P. and {Binet}, V. and {Bismark}, K. and {Bister}, T. and {Biteau}, J. and {Blazek}, J. and {Bleve}, C. and {Boh{\'a}{\v{c}}ov{\'a}}, M. and {Boncioli}, D. and {Bonifazi}, C. and {Bonneau Arbeletche}, L. and {Borodai}, N. and {Botti}, A.~M. and {Brack}, J. and {Bretz}, T. and {Brichetto Orchera}, P.~G. and {Briechle}, F.~L. and {Buchholz}, P. and {Bueno}, A. and {Buitink}, S. and {Buscemi}, M. and {B{\"u}sken}, M. and {Caballero-Mora}, K.~S. and {Caccianiga}, L. and {Canfora}, F. and {Caracas}, I. and {Carceller}, J.~M. and {Caruso}, R. and {Castellina}, A. and {Catalani}, F. and {Cataldi}, G. and {Cazon}, L. and {Cerda}, M. and {Chinellato}, J.~A. and {Chudoba}, J. and {Chytka}, L. and {Clay}, R.~W. and {Cobos Cerutti}, A.~C. and {Colalillo}, R. and {Coleman}, A. and {Coluccia}, M.~R. and {Concei{\c{c}}{\~a}o}, R. and {Condorelli}, A. and {Consolati}, G. and {Contreras}, F. and {Convenga}, F. and {Correia dos Santos}, D. and {Covault}, C.~E. and {Dasso}, S. and {Daumiller}, K. and {Dawson}, B.~R. and {Day}, J.~A. and {de Almeida}, R.~M. and {de Jes{\'u}s}, J. and {de Jong}, S.~J. and {De Mauro}, G. and {de Mello Neto}, J.~R.~T. and {De Mitri}, I. and {de Oliveira}, J. and {de Oliveira Franco}, D. and {de Palma}, F. and {de Souza}, V. and {De Vito}, E. and {del R{\'\i}o}, M. and {Deligny}, O. and {Di Matteo}, A. and {Dobrigkeit}, C. and {D'Olivo}, J.~C. and {Domingues Mendes}, L.~M. and {dos Anjos}, R.~C. and {dos Santos}, D. and {Dova}, M.~T. and {Ebr}, J. and {Engel}, R. and {Epicoco}, I. and {Erdmann}, M. and {Escobar}, C.~O. and {Etchegoyen}, A. and {Falcke}, H. and {Farmer}, J. and {Farrar}, G. and {Fauth}, A.~C. and {Fazzini}, N. and {Feldbusch}, F. and {Fenu}, F. and {Fick}, B. and {Figueira}, J.~M. and {Filip{\v{c}}i{\v{c}}}, A. and {Fitoussi}, T. and {Fodran}, T. and {Freire}, M.~M. and {Fujii}, T. and {Fuster}, A. and {Galea}, C. and {Galelli}, C. and {Garc{\'\i}a}, B. and {Garcia Vegas}, A.~L. and {Gemmeke}, H. and {Gesualdi}, F. and {Gherghel-Lascu}, A. and {Ghia}, P.~L. and {Giaccari}, U. and {Giammarchi}, M. and {Glombitza}, J. and {Gobbi}, F. and {Gollan}, F. and {Golup}, G. and {G{\'o}mez Berisso}, M. and {G{\'o}mez Vitale}, P.~F. and {Gongora}, J.~P. and {Gonz{\'a}lez}, J.~M. and {Gonz{\'a}lez}, N. and {Goos}, I. and {G{\'o}ra}, D. and {Gorgi}, A. and {Gottowik}, M. and {Grubb}, T.~D. and {Guarino}, F. and {Guedes}, G.~P. and {Guido}, E. and {Hahn}, S. and {Hamal}, P. and {Hampel}, M.~R. and {Hansen}, P. and {Harari}, D. and {Harvey}, V.~M. and {Haungs}, A. and {Hebbeker}, T. and {Heck}, D. and {Hill}, G.~C. and {Hojvat}, C. and {H{\"o}randel}, J.~R. and {Horvath}, P. and {Hrabovsk{\'y}}, M. and {Huege}, T. and {Insolia}, A. and {Isar}, P.~G. and {Janecek}, P. and {Johnsen}, J.~A. and {Jurysek}, J. and {K{\"a}{\"a}p{\"a}}, A. and {Kampert}, K.~H. and {Karastathis}, N. and {Keilhauer}, B. and {Kemp}, J. and {Khakurdikar}, A. and {Kizakke Covilakam}, V.~V. and {Klages}, H.~O. and {Kleifges}, M. and {Kleinfeller}, J. and {K{\"o}pke}, M. and {Kunka}, N. and {Lago}, B.~L. and {Lang}, R.~G. and {Langner}, N. and {Leigui de Oliveira}, M.~A. and {Lenok}, V. and {Letessier-Selvon}, A. and {Lhenry-Yvon}, I. and {Lo Presti}, D. and {Lopes}, L. and {L{\'o}pez}, R. and {Lu}, L. and {Luce}, Q. and {Lundquist}, J.~P. and {Machado Payeras}, A. and {Mancarella}, G. and {Mandat}, D. and {Manning}, B.~C. and {Manshanden}, J. and {Mantsch}, P. and {Marafico}, S.},
        title = "{The energy spectrum of cosmic rays beyond the turn-down around {}10$^{17}$ eV as measured with the surface detector of the Pierre Auger Observatory}",
      journal = {European Physical Journal C},
         year = 2021,
        month = nov,
       volume = {81},
       number = {11},
          eid = {966},
        pages = {966},
          doi = {10.1140/epjc/s10052-021-09700-w},
archivePrefix = {arXiv},
       eprint = {2109.13400},
 primaryClass = {astro-ph.HE},
       adsurl = {https://ui.adsabs.harvard.edu/abs/2021EPJC...81..966A}
}

@ARTICLE{DeRujula-2019,
       author = {{De R{\'u}jula}, A.},
        title = "{The cosmic-ray spectra: News on their knees}",
      journal = {Physics Letters B},
         year = 2019,
        month = mar,
       volume = {790},
        pages = {444-452},
          doi = {10.1016/j.physletb.2019.01.059},
       adsurl = {https://ui.adsabs.harvard.edu/abs/2019PhLB..790..444D}
}

@ARTICLE{Recchia-Gabici-2024,
       author = {{Recchia}, S. and {Gabici}, S.},
        title = "{Origin of the spectral features observed in the cosmic-ray spectrum}",
      journal = {\aap},
         year = 2024,
        month = dec,
       volume = {692},
          eid = {A20},
        pages = {A20},
          doi = {10.1051/0004-6361/202349005},
archivePrefix = {arXiv},
       eprint = {2312.11397},
 primaryClass = {astro-ph.HE},
       adsurl = {https://ui.adsabs.harvard.edu/abs/2024A&A...692A..20R}
}

@ARTICLE{Aguilar-etal-2015,
       author = {{Aguilar}, M. and {Aisa}, D. and {Alpat}, B. and {Alvino}, A. and {Ambrosi}, G. and {Andeen}, K. and {Arruda}, L. and {Attig}, N. and {Azzarello}, P. and {Bachlechner}, A. and {Barao}, F. and {Barrau}, A. and {Barrin}, L. and {Bartoloni}, A. and {Basara}, L. and {Battarbee}, M. and {Battiston}, R. and {Bazo}, J. and {Becker}, U. and {Behlmann}, M. and {Beischer}, B. and {Berdugo}, J. and {Bertucci}, B. and {Bigongiari}, G. and {Bindi}, V. and {Bizzaglia}, S. and {Bizzarri}, M. and {Boella}, G. and {de Boer}, W. and {Bollweg}, K. and {Bonnivard}, V. and {Borgia}, B. and {Borsini}, S. and {Boschini}, M.~J. and {Bourquin}, M. and {Burger}, J. and {Cadoux}, F. and {Cai}, X.~D. and {Capell}, M. and {Caroff}, S. and {Casaus}, J. and {Cascioli}, V. and {Castellini}, G. and {Cernuda}, I. and {Cerreta}, D. and {Cervelli}, F. and {Chae}, M.~J. and {Chang}, Y.~H. and {Chen}, A.~I. and {Chen}, H. and {Cheng}, G.~M. and {Chen}, H.~S. and {Cheng}, L. and {Chou}, H.~Y. and {Choumilov}, E. and {Choutko}, V. and {Chung}, C.~H. and {Clark}, C. and {Clavero}, R. and {Coignet}, G. and {Consolandi}, C. and {Contin}, A. and {Corti}, C. and {Gil}, E. Cortina and {Coste}, B. and {Creus}, W. and {Crispoltoni}, M. and {Cui}, Z. and {Dai}, Y.~M. and {Delgado}, C. and {Della Torre}, S. and {Demirk{\"o}z}, M.~B. and {Derome}, L. and {Di Falco}, S. and {Di Masso}, L. and {Dimiccoli}, F. and {D{\'\i}az}, C. and {von Doetinchem}, P. and {Donnini}, F. and {Du}, W.~J. and {Duranti}, M. and {D'Urso}, D. and {Eline}, A. and {Eppling}, F.~J. and {Eronen}, T. and {Fan}, Y.~Y. and {Farnesini}, L. and {Feng}, J. and {Fiandrini}, E. and {Fiasson}, A. and {Finch}, E. and {Fisher}, P. and {Galaktionov}, Y. and {Gallucci}, G. and {Garc{\'\i}a}, B. and {Garc{\'\i}a-L{\'o}pez}, R. and {Gargiulo}, C. and {Gast}, H. and {Gebauer}, I. and {Gervasi}, M. and {Ghelfi}, A. and {Gillard}, W. and {Giovacchini}, F. and {Goglov}, P. and {Gong}, J. and {Goy}, C. and {Grabski}, V. and {Grandi}, D. and {Graziani}, M. and {Guandalini}, C. and {Guerri}, I. and {Guo}, K.~H. and {Haas}, D. and {Habiby}, M. and {Haino}, S. and {Han}, K.~C. and {He}, Z.~H. and {Heil}, M. and {Hoffman}, J. and {Hsieh}, T.~H. and {Huang}, Z.~C. and {Huh}, C. and {Incagli}, M. and {Ionica}, M. and {Jang}, W.~Y. and {Jinchi}, H. and {Kanishev}, K. and {Kim}, G.~N. and {Kim}, K.~S. and {Kirn}, Th. and {Kossakowski}, R. and {Kounina}, O. and {Kounine}, A. and {Koutsenko}, V. and {Krafczyk}, M.~S. and {La Vacca}, G. and {Laudi}, E. and {Laurenti}, G. and {Lazzizzera}, I. and {Lebedev}, A. and {Lee}, H.~T. and {Lee}, S.~C. and {Leluc}, C. and {Levi}, G. and {Li}, H.~L. and {Li}, J.~Q. and {Li}, Q. and {Li}, Q. and {Li}, T.~X. and {Li}, W. and {Li}, Y. and {Li}, Z.~H. and {Li}, Z.~Y. and {Lim}, S. and {Lin}, C.~H. and {Lipari}, P. and {Lippert}, T. and {Liu}, D. and {Liu}, H. and {Lolli}, M. and {Lomtadze}, T. and {Lu}, M.~J. and {Lu}, S.~Q. and {Lu}, Y.~S. and {Luebelsmeyer}, K. and {Luo}, J.~Z. and {Lv}, S.~S. and {Majka}, R. and {Ma{\~n}{\'a}}, C. and {Mar{\'\i}n}, J. and {Martin}, T. and {Mart{\'\i}nez}, G. and {Masi}, N. and {Maurin}, D. and {Menchaca-Rocha}, A. and {Meng}, Q. and {Mo}, D.~C. and {Morescalchi}, L. and {Mott}, P. and {M{\"u}ller}, M. and {Ni}, J.~Q. and {Nikonov}, N. and {Nozzoli}, F. and {Nunes}, P. and {Obermeier}, A. and {Oliva}, A. and {Orcinha}, M. and {Palmonari}, F. and {Palomares}, C. and {Paniccia}, M. and {Papi}, A. and {Pauluzzi}, M. and {Pedreschi}, E. and {Pensotti}, S. and {Pereira}, R. and {Picot-Clemente}, N. and {Pilo}, F. and {Piluso}, A. and {Pizzolotto}, C. and {Plyaskin}, V.},
        title = "{Precision Measurement of the Proton Flux in Primary Cosmic Rays from Rigidity 1 GV to 1.8 TV with the Alpha Magnetic Spectrometer on the International Space Station}",
      journal = {\prl},
         year = 2015,
        month = may,
       volume = {114},
       number = {17},
          eid = {171103},
        pages = {171103},
          doi = {10.1103/PhysRevLett.114.171103},
       adsurl = {https://ui.adsabs.harvard.edu/abs/2015PhRvL.114q1103A}
}

@ARTICLE{Adriani-etal-2023,
       author = {{Adriani}, O. and {Akaike}, Y. and {Asano}, K. and {Asaoka}, Y. and {Berti}, E. and {Bigongiari}, G. and {Binns}, W.~R. and {Bongi}, M. and {Brogi}, P. and {Bruno}, A. and {Buckley}, J.~H. and {Cannady}, N. and {Castellini}, G. and {Checchia}, C. and {Cherry}, M.~L. and {Collazuol}, G. and {de Nolfo}, G.~A. and {Ebisawa}, K. and {Ficklin}, A.~W. and {Fuke}, H. and {Gonzi}, S. and {Guzik}, T.~G. and {Hams}, T. and {Hibino}, K. and {Ichimura}, M. and {Ioka}, K. and {Ishizaki}, W. and {Israel}, M.~H. and {Kasahara}, K. and {Kataoka}, J. and {Kataoka}, R. and {Katayose}, Y. and {Kato}, C. and {Kawanaka}, N. and {Kawakubo}, Y. and {Kobayashi}, K. and {Kohri}, K. and {Krawczynski}, H.~S. and {Krizmanic}, J.~F. and {Maestro}, P. and {Marrocchesi}, P.~S. and {Messineo}, A.~M. and {Mitchell}, J.~W. and {Miyake}, S. and {Moiseev}, A.~A. and {Mori}, M. and {Mori}, N. and {Motz}, H.~M. and {Munakata}, K. and {Nakahira}, S. and {Nishimura}, J. and {Okuno}, S. and {Ormes}, J.~F. and {Ozawa}, S. and {Pacini}, L. and {Papini}, P. and {Rauch}, B.~F. and {Ricciarini}, S.~B. and {Sakai}, K. and {Sakamoto}, T. and {Sasaki}, M. and {Shimizu}, Y. and {Shiomi}, A. and {Spillantini}, P. and {Stolzi}, F. and {Sugita}, S. and {Sulaj}, A. and {Takita}, M. and {Tamura}, T. and {Terasawa}, T. and {Torii}, S. and {Tsunesada}, Y. and {Uchihori}, Y. and {Vannuccini}, E. and {Wefel}, J.~P. and {Yamaoka}, K. and {Yanagita}, S. and {Yoshida}, A. and {Yoshida}, K. and {Zober}, W.~V. and {Calet Collaboration}},
        title = "{Direct Measurement of the Cosmic-Ray Helium Spectrum from 40 GeV to 250 TeV with the Calorimetric Electron Telescope on the International Space Station}",
      journal = {\prl},
         year = 2023,
        month = apr,
       volume = {130},
       number = {17},
          eid = {171002},
        pages = {171002},
          doi = {10.1103/PhysRevLett.130.171002},
archivePrefix = {arXiv},
       eprint = {2304.14699},
 primaryClass = {astro-ph.HE},
       adsurl = {https://ui.adsabs.harvard.edu/abs/2023PhRvL.130q1002A}
}

@ARTICLE{Bhadra-etal-2025,
       author = {{Bhadra}, Sourav and {Thoudam}, Satyendra and {Nath}, Biman B. and {Sharma}, Prateek},
        title = "{The TeV Spectral Bump of Cosmic-Ray Protons and Helium Nuclei: The Role of Nearby Supernova Remnants}",
      journal = {\apj},
         year = 2025,
        month = aug,
       volume = {989},
       number = {1},
          eid = {74},
        pages = {74},
          doi = {10.3847/1538-4357/ade796},
archivePrefix = {arXiv},
       eprint = {2506.18681},
 primaryClass = {astro-ph.HE},
       adsurl = {https://ui.adsabs.harvard.edu/abs/2025ApJ...989...74B}
}

@ARTICLE{Biermann-etal-2010,
       author = {{Biermann}, Peter L. and {Becker}, Julia K. and {Dreyer}, Jens and {Meli}, Athina and {Seo}, Eun-Suk and {Stanev}, Todor},
        title = "{The Origin of Cosmic Rays: Explosions of Massive Stars with Magnetic Winds and Their Supernova Mechanism}",
      journal = {\apj},
         year = 2010,
        month = dec,
       volume = {725},
       number = {1},
        pages = {184-187},
          doi = {10.1088/0004-637X/725/1/184},
archivePrefix = {arXiv},
       eprint = {1009.5592},
 primaryClass = {astro-ph.HE},
       adsurl = {https://ui.adsabs.harvard.edu/abs/2010ApJ...725..184B}
}

@ARTICLE{Qian-etal-2025,
       author = {{Qian}, Xiang-Li and {Sun}, Hui-Ying and {Nie}, Lin and {Ge}, Yu-Hai and {Liu}, Wei and {Guo}, Yi-Qing},
        title = "{Monogem's Contribution to the Galactic Cosmic-Ray Spectrum}",
      journal = {\apj},
         year = 2025,
        month = aug,
       volume = {989},
       number = {1},
          eid = {96},
        pages = {96},
          doi = {10.3847/1538-4357/adef51},
       adsurl = {https://ui.adsabs.harvard.edu/abs/2025ApJ...989...96Q}
}

@article{Beresnyak2011,
  title = {Spectral Slope and Kolmogorov Constant of MHD Turbulence},
  author = {Beresnyak, A.},
  journal = {Phys. Rev. Lett.},
  volume = {106},
  issue = {7},
  pages = {075001},
  numpages = {4},
  year = {2011},
  month = {Feb},
  publisher = {American Physical Society},
  doi = {10.1103/PhysRevLett.106.075001},
  url = {https://link.aps.org/doi/10.1103/PhysRevLett.106.075001}
}

@ARTICLE{Verma1996,
       author = {{Verma}, M.~K. and {Roberts}, D.~A. and {Goldstein}, M.~L. and {Ghosh}, S. and {Stribling}, W.~T.},
        title = "{A numerical study of the nonlinear cascade of energy in magnetohydrodynamic turbulence}",
      journal = {\jgr},
         year = 1996,
        month = oct,
       volume = {101},
       number = {A10},
        pages = {21619-21626},
          doi = {10.1029/96JA01773},
       adsurl = {https://ui.adsabs.harvard.edu/abs/1996JGR...10121619V}
}

@ARTICLE{Gleeson1968,
       author = {{Gleeson}, L.~J.},
        title = "{Emerging Theories of the Solar Modulation of Cosmic Rays}",
      journal = {\pasa},
         year = 1968,
        month = dec,
       volume = {1},
       number = {4},
        pages = {130-132},
          doi = {10.1017/S1323358000011036},
       adsurl = {https://ui.adsabs.harvard.edu/abs/1968PASA....1..130G}
}

@article{Haverkorn_2008,
doi = {10.1086/587165},
url = {https://doi.org/10.1086/587165},
year = {2008},
month = {jun},
publisher = {},
volume = {680},
number = {1},
pages = {362},
author = {Haverkorn, M. and Brown, J. C. and Gaensler, B. M. and McClure-Griffiths, N. M.},
title = {The Outer Scale of Turbulence in the Magnetoionized Galactic Interstellar Medium},
journal = {The Astrophysical Journal}
}

@article{Evoli2018,
  title = {Origin of the Cosmic Ray Galactic Halo Driven by Advected Turbulence and Self-Generated Waves},
  author = {Evoli, Carmelo and Blasi, Pasquale and Morlino, Giovanni and Aloisio, Roberto},
  journal = {Phys. Rev. Lett.},
  volume = {121},
  issue = {2},
  pages = {021102},
  numpages = {5},
  year = {2018},
  month = {Jul},
  publisher = {American Physical Society},
  doi = {10.1103/PhysRevLett.121.021102},
  url = {https://link.aps.org/doi/10.1103/PhysRevLett.121.021102}
}

@article{Yeung1997,
  title = {Universality of the Kolmogorov constant in numerical simulations of turbulence},
  author = {Yeung, P. K. and Zhou, Ye},
  journal = {Phys. Rev. E},
  volume = {56},
  issue = {2},
  pages = {1746--1752},
  numpages = {0},
  year = {1997},
  month = {Aug},
  publisher = {American Physical Society},
  doi = {10.1103/PhysRevE.56.1746},
  url = {https://link.aps.org/doi/10.1103/PhysRevE.56.1746}
}

@article{Sreenivasan1995,
    author = {Sreenivasan, Katepalli R.},
    title = {On the universality of the Kolmogorov constant},
    journal = {Physics of Fluids},
    volume = {7},
    number = {11},
    pages = {2778-2784},
    year = {1995},
    month = {11},
    issn = {1070-6631},
    doi = {10.1063/1.868656},
    url = {https://doi.org/10.1063/1.868656},
    eprint = {https://pubs.aip.org/aip/pof/article-pdf/7/11/2778/19257722/2778_1_online.pdf},
}

@article{Biskamp1989,
    author = {Biskamp, D. and Welter, H.},
    title = {Dynamics of decaying two‐dimensional magnetohydrodynamic turbulence},
    journal = {Physics of Fluids B: Plasma Physics},
    volume = {1},
    number = {10},
    pages = {1964-1979},
    year = {1989},
    month = {10},
    issn = {0899-8221},
    doi = {10.1063/1.859060},
    url = {https://doi.org/10.1063/1.859060},
    eprint = {https://pubs.aip.org/aip/pfb/article-pdf/1/10/1964/12506975/1964_1_online.pdf},
}

@ARTICLE{MatthaeusZhou1989,
       author = {{Matthaeus}, William H. and {Zhou}, Ye},
        title = "{Extended inertial range phenomenology of magnetohydrodynamic turbulence}",
      journal = {Physics of Fluids B},
         year = 1989,
        month = sep,
       volume = {1},
       number = {9},
        pages = {1929-1931},
          doi = {10.1063/1.859110},
       adsurl = {https://ui.adsabs.harvard.edu/abs/1989PhFlB...1.1929M}
}
\bibliographystyle{aasjournal}
\end{document}